\documentclass{customjournal}
\usepackage{todonotes} 
\usepackage{cite}
\usepackage{xurl}
\usepackage{amsmath,amssymb,amsfonts}
\usepackage{algorithmic}
\usepackage{graphicx}
\usepackage{subcaption}
\usepackage{indentfirst}
\usepackage{textcomp}
\usepackage{xcolor}
\usepackage{hyperref}
\usepackage{float}
\usepackage{cuted}
\usepackage{capt-of}
\usepackage{placeins}
\usepackage{booktabs} 
\usepackage{siunitx}
\hypersetup{
    colorlinks=true,
    linkcolor=blue,
    filecolor=magenta,
    urlcolor=black,
    citecolor=blue
}

\def\BibTeX{{\rm B\kern-.05em{\sc i\kern-.025em b}\kern-.08em
    T\kern-.1667em\lower.7ex\hbox{E}\kern-.125emX}}
\begin{document}
\pagestyle{plain} 


\title{Smart Navigation for Visual Prostheses in Virtual Reality: An End-to-End Framework for Priority-Based Scene Translation and Path Guidance}


\author[1]{Mohamed H. Abdellatif$^{*}$}
\author[1]{Fatma S. Elsharkawy$^{*}$}
\author[1]{Nouran H. Qassem$^{*}$}
\author[1]{Talal M. Emara}
\author[1]{Nada K. Kotb}
\author[1,2]{\href{mailto:mrushdi@eng1.cu.edu.eg}{Muhammad Rushdi}}
\author[2]{Reham H. Elnabawy}

\renewcommand{\thefootnote}{*}
\footnotetext{These authors contributed equally to this work.}

\affil[1]{Department of Biomedical Engineering and Systems, Faculty of Engineering, Cairo University, Giza, Egypt}
\affil[2]{School of Information Technology, New Giza University, Giza, Egypt}

\thispagestyle{plain} 

\maketitle

\begin{abstract}
Visual prosthetics provide a promising direction for partial restoration of functional vision for people with total retinal blindness. However, existing systems face significant challenges in translating complex visual scenes into meaningful perceptions due to limited spatial resolution, leading to difficulties in scene understanding. Furthermore, existing solutions don't adequately account for user requirements and concerns, and this creates a significant gap between user expectations and the developed solutions. To address these gaps, we conducted interviews with 10 blind human subjects. These interviews essentially indicated that the main key challenge for these people is outdoor navigation. In this paper, we present an end-to-end smart navigation system for visual prostheses users. Our approach employs a three-stage framework. First, an object detector is implemented to identify and localize points
of interest for navigation tasks in pedestrian environments, and simultaneously generate the safest paths available to the users. Second, the detected objects are abstracted into simple geometric shapes suitable for low-spatial-resolution vision. The detected objects are filtered based on a multi-criteria priority scoring function. Finally, this information is encoded into optimized stimulation parameters, which are fed into the visual prosthesis implants to generate enhanced phosphene representations for obstacle avoidance and path planning. This whole system is validated with sighted participants using virtual reality to simulate outdoor navigation. Our smart navigation system improves user independence while taking into account the limitations of visual prosthetics.
\end{abstract}



\section{Introduction}
Globally, blindness and visual impairment affect approximately 40 million and 250 million people, respectively, severely hindering independence and diminishing overall quality of life \cite{WHO2023}. Individuals with total retinal blindness, such as those suffering from retinitis pigmentosa (RP) or age-related macular degeneration (AMD), still retain intact optic nerves and visual cortices. Thus, such patients can benefit from visual prosthetics, particularly retinal implants \cite{Shim2020RetinalProsthetic}\cite{YUE201621}. However, current neuroprosthetic technology remains limited in its ability to restore functional, task-oriented vision \cite{Fernandez2018VisualNeuroprostheses}.\\
The human visual system is a highly organized, hierarchical pathway. After light enters through the cornea and lens and is focused on the retina, photoreceptors (rods and cones) convert this light into neural signals. These signals propagate through retinal ganglion cells to the optic nerve, relay through the lateral geniculate nucleus (LGN) of the thalamus, and ultimately arrive in the primary visual cortex (V1) \cite{Wandell1995FoundationsVision}. Patients with retinitis pigmentosa (RP) or age-related macular degeneration (AMD) suffer damage at different points along this pathway, necessitating different interventions. In cases where the photoreceptors are severely damaged but the higher neural structures remain functional, retinal implants are the most viable prosthetic approach.\\
Visual prosthetics operate by electrically stimulating surviving neurons along this pathway—at the retinal, thalamic, or cortical levels \cite{Maghami2014VisualProstheses}. This electrical stimulation evokes localized perceptions of light known as phosphenes \cite{CHEN20091493}. Ideally, phosphenes appear in the user’s visual field as discrete points of light, elicited via electrode arrays that deliver controlled electrical pulses. However, because of the underlying biology of the retina and ganglion cell axons, the phosphenes often appear not as perfect points of light but as elongated streaks or tails. The axon map model simulates the shape of the axons in the visual pathway to reproduce these elongated phosphene percepts \cite{beyeler2019retinal}. When multiple electrodes are activated simultaneously, phosphene patterns form basic shapes \cite{Chen2020ShapePerception}. The perceptual quality of these shapes is fundamentally bottlenecked by the low spatial and temporal resolution of current electrode arrays\cite{kucukoglu2025} \cite{Fernandez2018VisualNeuroprostheses}.\\ Despite this limitation, existing research in visual prosthetics has generally followed two approaches. The first and most traditional approach attempts to directly reproduce raw visual scenes (direct mapping). However, this is obviously impractical; given the low resolution of implants, direct mapping often results in severe visual clutter rather than meaningful scene comprehension. A second, more recent approach has recognized this limited bandwidth, and tried various techniques in image processing and computer vision to selectively filter the environment and condense the information presented to the prosthetic user (a review of these approaches is elaborated upon in Section \ref{subsec:lit_review}). While these selective encoding methods represent a significant step forward, they often lack a structured focus on what information is actually most important for a visually impaired user navigating complex environments. \\ Furthermore, a critical lack of user-centered design has historically limited the adoption and usability of visual prosthetics \cite{nadolskis2024} \cite{karadima2023}. \\



\begin{figure*}[htbp]
    \centering
    \includegraphics[width=\textwidth]{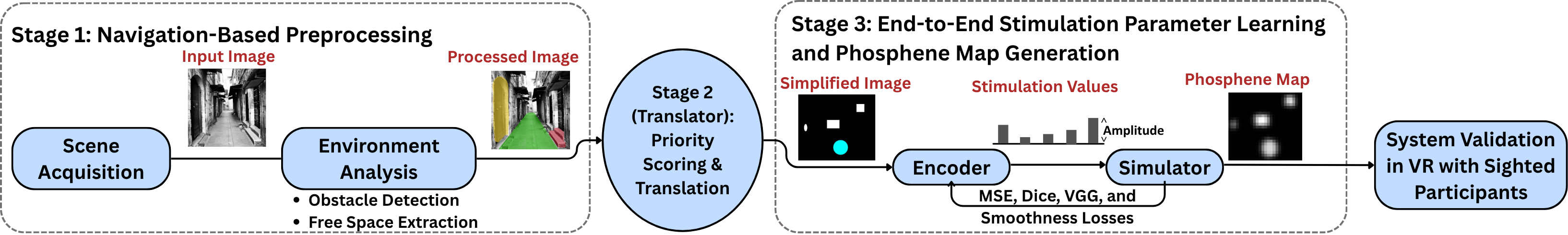} 
    \caption{The detailed system pipeline showing the end-to-end data flow from scene acquisition to phosphene map generation.}
    \label{fig:pipeline}
\end{figure*}

In this paper, we address the compounding challenges of prosthetic vision by proposing a user-centered end-to-end visual prosthetic framework, validated in a virtual reality simulation. The pipeline is tailored to the PRIMA subretinal implant and restricted to its hardware limitations: primarily its limited $7^\circ$ degree field of view (FOV)  and its limited resolution (378 electrodes). The proposed approach can still be extended to other prosthetics as well. The key contributions of this article are of four folds: \\
\textbf{First}, we conducted a user-centered survey to identify the needs of visually impaired people and what tasks they find difficult to do. Their needs and thoughts about what matters to them guided the design of our framework. \\
\textbf{Second}, to the best of our knowledge, no earlier work addressed the design of a task-specific visual prosthetic solution for outdoor navigation. \\ \textbf{Third}, we introduced a ``priority scoring and translation module", as shown in Fig. \ref{fig:pipeline}, to translate complex visual scenes into simpler, priority-based language. This design aligns more effectively with the constraints of limited-resolution prosthetics.\\ \textbf{Fourth}, the proposed framework was validated using `AfterDawn'*, 
an open-source, custom-built virtual reality (VR) framework that enables 
researchers to develop VR experiments and scenarios in an easy and flexible 
manner. It also provides a toolkit for recording and measuring navigation 
metrics during VR experiments. Sighted volunteers used the environment to 
evaluate the proposed system in real time.

\footnotetext{*The source code for AfterDawn is publicly available at \\
\url{https://github.com/TalalEmara/Dawn-Pilot--AfterDawn}.}

\subsection{Related Work}
\label{subsec:lit_review}

\subsubsection{Traditional Scene-to-Phosphene Encoding} Visual prosthetic systems often rely on direct scene-to-phosphene mapping, either by mapping the raw scene\cite{BaylerExperiment}\cite{kucukoglu2025}, or by simplifying the scene and extracting the edges only to be mapped\cite{simpleNavigation}. However, due to the limited spatial resolution and narrow field of view of current retinal implants, rendering these scenes without targeted processing results in visual overcrowding or non-meaningful dots of phosphenes. Studies evaluating navigation tasks have shown that direct resizing of images to match electrode arrays fails to provide meaningful spatial comprehension, leading to high cognitive load and exceptionally poor wayfinding performance \cite{simpleNavigation}. Consequently, to facilitate effective mobility, it is essential to highlight pertinent information within the surroundings rather than attempting raw scene reproduction \cite{BaylerExperiment}.

\subsubsection{Advanced processing for better scene representation} To address visual clutter, recent assistive frameworks have integrated deep learning models to filter scenes. Early approaches, such as the work by Sanchez-Garcia et al. \cite{sanchezgarcia2020}, utilized fully convolutional networks to extract structural lines and semantic objects for indoor environments. Additionally, Han et al. \cite{han2021} demonstrated that semantic object segmentation provides superior scene understanding for bionic vision compared to raw depth or visual saliency. However, their validation relied on pre-rendered offline videos, and they noted that naively combining spatial and semantic features actually degraded user performance due to information overload \cite{han2021}.

\subsubsection{Deep Learning for Stimulation Optimization} 
Beyond scene filtering, optimizing the actual electrical stimulation parameters is critical for generating intelligible phosphenes. Conventional systems use open-loop encoding, linearly mapping pixels to amplitudes without considering complex neurophysiological distortions such as the axonal effect \cite{beyeler2019retinal}. Recent advancements have introduced deep stimulus encoding, in which deep learning models are trained to optimize stimulation strategies based on patient-specific variations, using human-in-the-loop Bayesian optimization\cite{granley2023}.
A significant step forward is the end-to-end optimization paradigm, where a differentiable phosphene simulator acts as a fixed decoder, enabling gradient-based learning of stimulation parameters tied directly to perceptual outcomes. Two notable implementations of this idea target different levels of the visual pathway. Granley et al.\cite{granley2022} applied this to retinal prostheses via the hybrid neural autoencoder (HNA), training a deep neural encoder to invert a differentiable forward model substantially outperforming naïve pixel-to-amplitude mappings. In parallel, van der Grinten et al.\cite{vandergrinten2024} developed an analogous framework for cortical prostheses, introducing a fully differentiable PyTorch-based simulator for cortical modeling. These methods establish encoder-decoder differentiable optimization architectures as a general and effective strategy for stimulus encoding, applicable across cortical neuroprosthetic architectures.
Building on this, K\"u\c{c}\"uko\u{g}lu et al. extended this approach to jointly optimize amplitude, pulse width, and frequency within biologically safe bounds using 3D convolutional and recurrent encoders and decoders trained on naturalistic navigation videos\cite{kucukoglu2025}.
Furthermore, optimizing the temporal and spatial scheduling of electrode activation (rastering) has been explored. Recent immersive simulated prosthetic vision (SPV) studies confirm that structured activation, using for example checkerboard patterns, significantly reduces motion artifacts and enhances perceptual clarity without increasing computational load \cite{kasowski2025}.

\subsubsection{VR Simulation Validation}
Validating advanced visual encoding schemes mostly relies on SPVs. Recent immersive VR studies have successfully demonstrated the feasibility of simplified SPVs for indoor mobility, utilizing contour-based cues \cite{deruyter2022} and depth profiles \cite{rasla2022}. 
Recent studies reveal that utilizing head-mounted displays (HMDs) significantly improves behavioral performance in obstacle avoidance tasks by providing necessary vestibular and proprioceptive cues \cite{kasowski2022immersive}. Furthermore, VR wayfinding experiments demonstrate that simplifying the visual rendering process—via limiting the viewing distance or extracting structural wireframe edges—reduces visual overcrowding and enhances users' cognitive mapping of unknown environments \cite{BaylerExperiment}. Immersive simulations have also been extended beyond basic mobility. In particular, experiments utilizing augmented-reality (AR) glasses confirm that human subjects can successfully perform complex daily tasks, including sentence reading and face recognition, despite severe pixelation and reduced fields of view \cite{ho2019performance}.

\subsubsection{New Directions}
Recent advancements reframe visual prosthetics from merely attempting to replicate natural sight into intelligent, highly adaptive assistive tools. This paradigm shift merges the concept of  “smart" bionic eyes—which utilize AI to simplify environments into task-specific visual augmentations—with  “neuroadaptive extended reality (XR)”, a closed-loop framework that jointly adapts the device and the brain based on real-time user intent, context, and cognitive state \cite{beyeler2025bionic}. By prioritizing patient-centered design and delivering sparse, personalized cues rather than overwhelming the user with raw data, these integrated systems significantly reduce cognitive load and transform implants into responsive, user-centric perceptual interfaces \cite{beyeler2022towards}.



\section{Methodology}
\subsection{User-Centric Design Requirements}
\label{subsec:user_centric}
Most current visual prosthetic systems suffer from significant disparities between engineering-driven development and actual daily needs of human users \cite{nadolskis2024}. Furthermore, surveys of potential users indicate that perceived risks—such as cognitive overload and visual confusion—often outweigh perceived benefits of visual prosthetic devices \cite{karadima2023}.

To bridge this gap and align our solution with real users, we engaged directly with a community of visually impaired people at the ``Dialogue in the Dark" exhibition in Egypt. Over two days and across two different batches, we experienced a complete simulation of total blindness with a visually impaired human guide. This included navigating physical replicas of Egypt's famous landmarks and attempting daily life activities, such as making a drink.

Following this immersive experience, we conducted one-to-one interviews with nine of the visually impaired guides. Informed consent  was obtained from each participant to record and analyze interview responses. We conducted a thematic analysis of these interview following Braun and Clarke (2006) \cite{Braun2006ThematicAnalysis}, with additional guidance from Ahmed et al. (2025) \cite{Ahmed2025ThematicAnalysis}. The most pressing daily challenges for the users were collected and summarized in Table \ref{tab:did_stats}, where frequencies represent the number of times each theme was mentioned across participants. Navigation is reported as the most faced challenge, and this directed our focus towards the problem of outdoor navigation. It was also revealed that for mobility, users prioritize immediate physical safety over detailed scene comprehension. Indeed, the priority for a blind or visually impaired person is knowing if an obstacle is present, rather than caring about what the obstacle type is.



\begin{table}[t]
\centering
\caption{Challenges perceived by 8 participants with visual impairments.}
\begin{tabular}{lcc}
\hline
\textbf{Theme} & \textbf{Frequency} & \textbf{Percentage} \\
\hline
Navigation & 30 & 50\% \\
Reading \& Writing & 14 & 23.3\% \\
Community, Society \& Environment & 3 & 5\% \\
Crossing Roads & 3 & 5\% \\
Outdoor & 3 & 5\% \\
Dependency & 2 & 3.3\% \\
Mobility \& Orientation & 2 & 3.3\% \\
Work Opportunities & 2 & 3.3\% \\
Adaptation & 1 & 1\% \\
\hline
\end{tabular}
\label{tab:did_stats}
\end{table}

\subsection{ Smart Navigation Pipeline}

Based on the thematic analysis outcomes (Section \ref{subsec:user_centric}), we adopt a “smart navigation” perspective where the primary goal is safe movement rather than detailed scene reconstruction. In other words, a visual prosthetic in this setting is considered as an assistive technology rather than an imitation of natural vision. Instead of overcrowding the limited prosthetic display area with raw visual data or unresolved details of one object, we focus exclusively on displaying the closest obstacles and free paths. Therefore, the key design criteria become collision avoidance, cognitive simplification, and path guidance. We propose a smart navigation pipeline that accounts for these design criteria as illustrated in Fig. \ref{fig:pipeline}.

Collectively, the proposed pipeline takes real-world scenes captured by an RGB-D camera as input and returns phosphene maps. This pipeline shall help newly blind people navigate outdoor environments via interpreting and following simplified phosphene images.   

The full pipeline consists of three stages. The first stage involves navigation-based preprocessing (NBP), where RGB-D scenes undergo different preprocessing steps. Specifically, key semantic information is extracted including all scene objects and free paths.
In the second stage, each object is assigned a priority score based on its distance from the viewer, relative objective velocity, object type (e.g. car, tree, etc.), and potential risk. Then, free paths are then filtered to avoid collision with high-priority objects (or obstacles). This process translates real-world scenes into highly simplified ones showing only high-priority objects and the best free path. This translation process acts as the bridge between complex visual scenes and adequate prosthetics-compliant scenes. Finally, in the third stage, stimulation parameters and phosphene maps are subsequently generated in an end-to-end fashion. In particular, an encoder is trained to map the simplified shapes into vectors of safe stimulation parameters. A biological visual prosthetics model is then simulated to generate the response to the stimulation parameters and hence generate the predicted phosphene maps.



\subsubsection{Navigation-based preprocessing}
In this stage, the sidewalk within an input urban scene is localized and segmented, and the corresponding free path is extracted and highlighted for users to navigate their environment safely. As well, objects within a given urban scene are accurately localized and classified in real-time.

\begin{figure*}[t]
    \centering

    \begin{subfigure}[b]{0.32\textwidth}
        \centering
        \includegraphics[width=\textwidth]{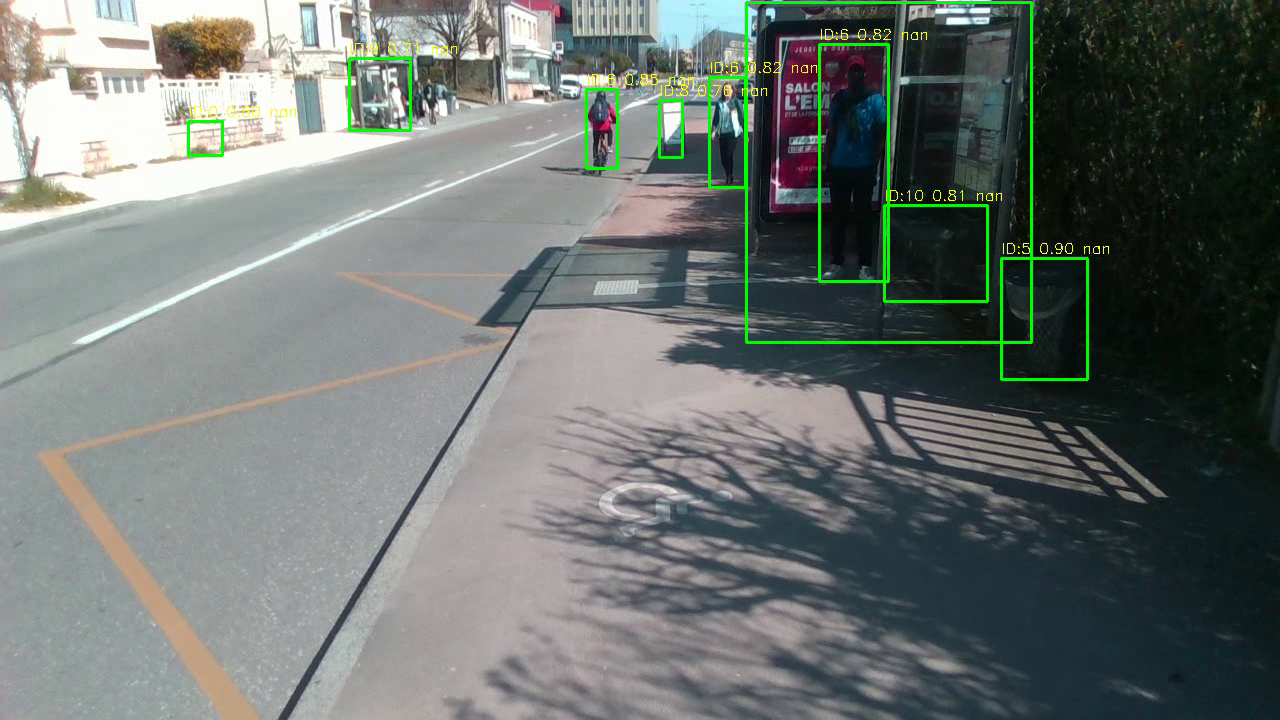}
        \caption{Input scene}
        \label{fig:sidewalk_input}
    \end{subfigure}
    \hfill
    \begin{subfigure}[b]{0.32\textwidth}
        \centering
        \includegraphics[width=\textwidth]{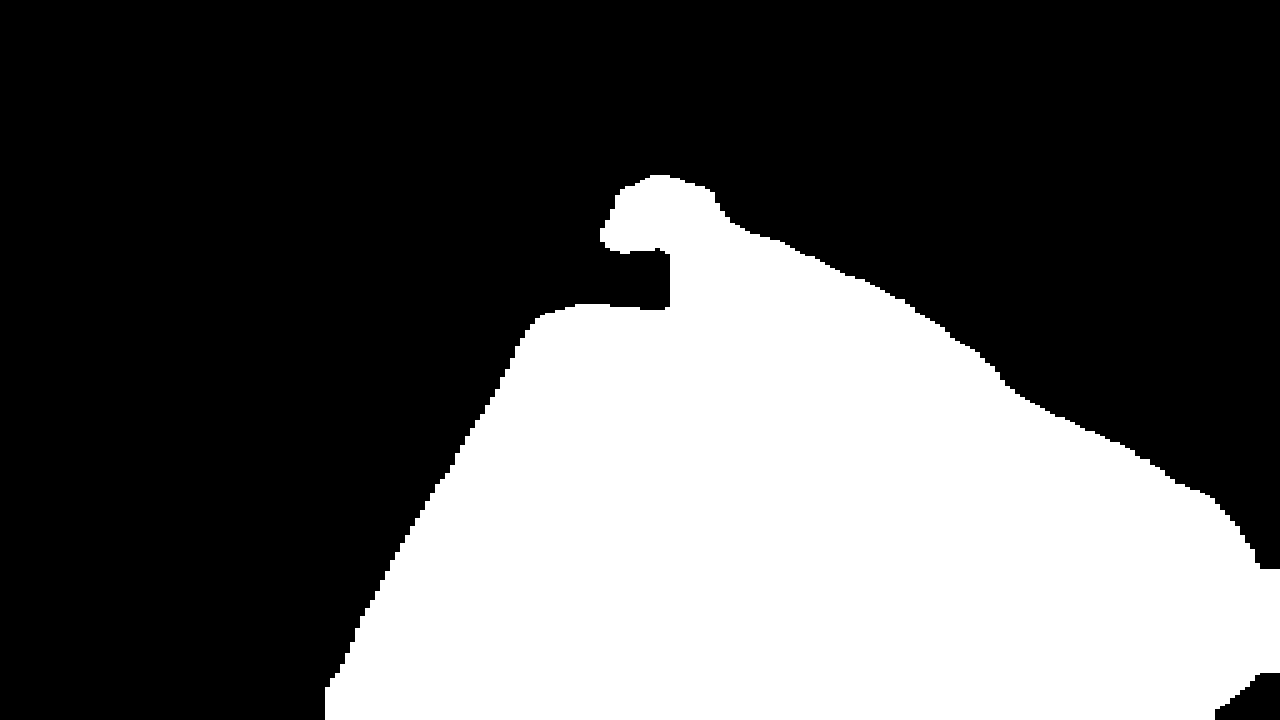}
        \caption{Free-path mask}
        \label{fig:sidewalk_segmentation}
    \end{subfigure}
    \hfill
    \begin{subfigure}[b]{0.32\textwidth}
        \centering
        \includegraphics[width=\textwidth]{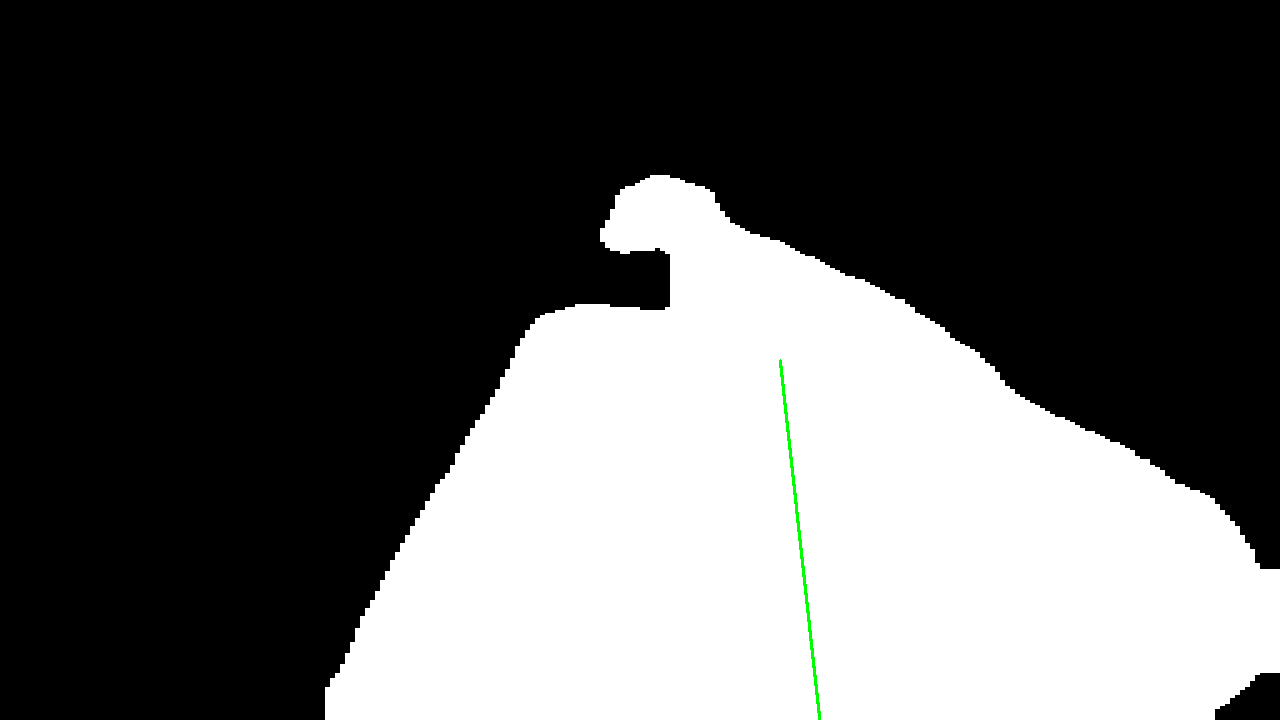}
        \caption{Extracted centerline}
        \label{fig:sidewalk_centerline}
    \end{subfigure}

    \vspace{0.5em}

    \begin{subfigure}[t]{0.42\textwidth}
        \centering
        \includegraphics[width=\textwidth]{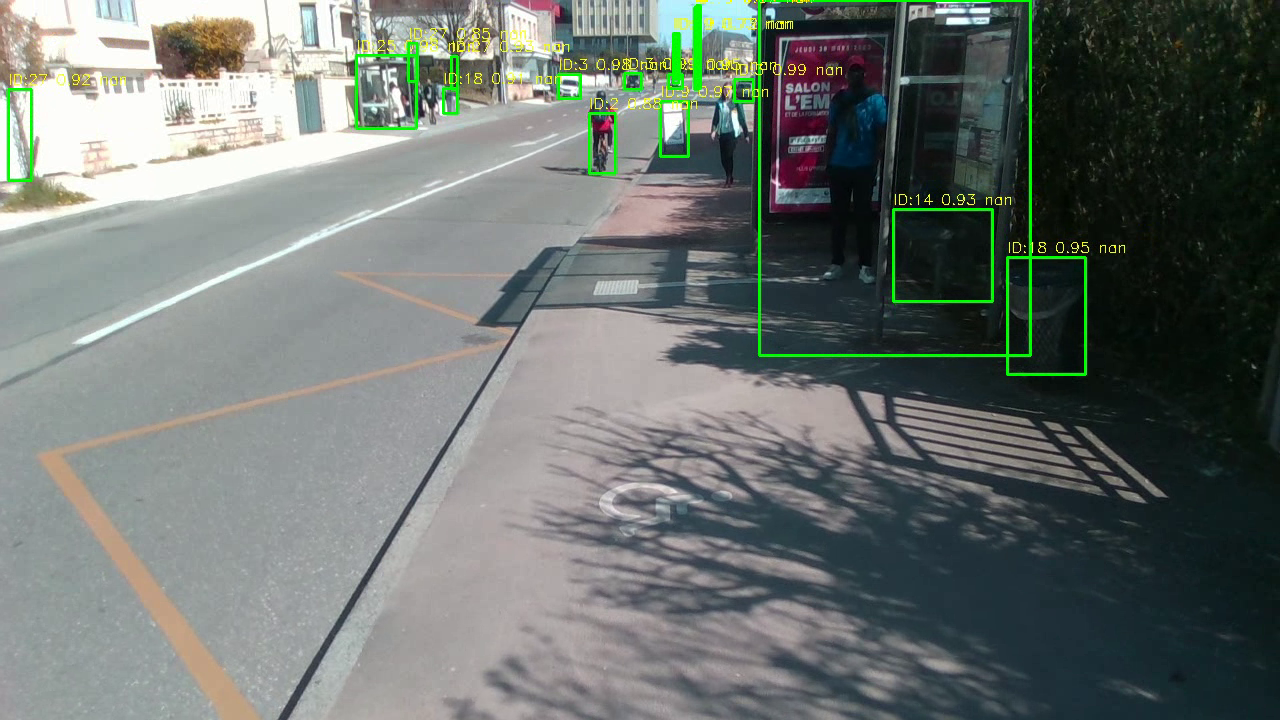}
        \caption{Faster R-CNN detections}
        \label{fig:rcnn_detection}
    \end{subfigure}
    \hfill
    \begin{subfigure}[t]{0.42\textwidth}
        \centering
        \includegraphics[width=\textwidth]{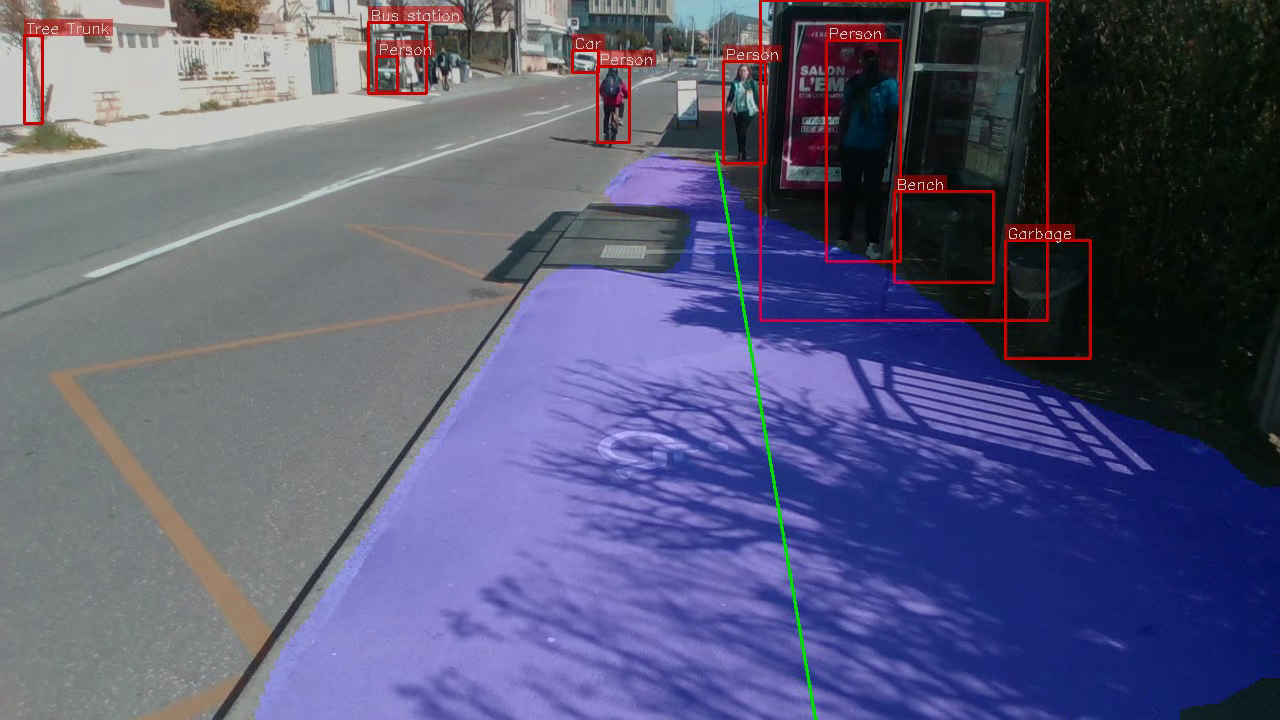}
        \caption{YOLOv8s detections with free-path segmentation and centerline}
        \label{fig:yolo_detection}
    \end{subfigure}

    \caption{Sidewalk segmentation, free-path centerline extraction, and
    obstacle detection. (a) Sample input scene. (b) Corresponding free-path
    mask obtained using the DeepLabv3 sidewalk segmentation model. (c) Extracted
    centerline of the free path, obtained by averaging the horizontal positions
    of the free-path mask across each image row and fitting a linear regression
    to the resulting positions. (d) Obstacle detections obtained using Faster
    R-CNN with a ResNet-50 backbone. (e) Obstacle detections obtained using
    YOLOv8s, shown together with the segmented free path and its extracted
    centerline.}
    \label{fig:sidewalk_pipeline}
\end{figure*}

 \paragraph{Sidewalk segmentation and free-path extraction} A DeepLabv3 \cite{deeplabv3} model trained on a dataset of 3000 images of urban footpaths in Bangladesh is used for sidewalk segmentation  \cite{afifaniks2023footpath}. The centerline of each free path is thus extracted via averaging the horizontal positions of the free-path mask across each image row, and applying a linear regression fit to the averaged positions. A sample input scene is shown in Fig. \ref{fig:sidewalk_pipeline}(a), with the corresponding sidewalk segmentation in Fig. \ref{fig:sidewalk_pipeline}(b) and extracted centerline in Fig. \ref{fig:sidewalk_pipeline}(c).

\paragraph{Obstacle detection}
To balance detection accuracy with the latency requirements of real-time operation, we considered two object detection architectures: Faster R-CNN with a ResNet-50 backbone and YOLOv8s. Sample outputs from the two detectors are shown in Fig. \ref{fig:sidewalk_pipeline}(d) and (e), respectively. Fig. \ref{fig:sidewalk_pipeline}(d) shows the obstacle detections produced by Faster R-CNN, while Fig. \ref{fig:sidewalk_pipeline}(e) shows the YOLOv8s detections together with the segmented free path and its extracted centerline.

    


The obstacle detection models are trained and evaluated on the combination of two datasets: the UBViseoGeoloc dataset \cite{ubvisiogeoloc}, and a synthetic dataset. The UBViseoGeoloc public dataset provides RGB-D sequences with bounding box annotations for objects relevant to pedestrian mobility (e.g., people, bus stations, trees) \cite{ubvisiogeoloc}. Since this dataset did not contain sufficient synthetic scenes, a supplementary dataset of 300 scenes was created using the A-Frame framework \cite{aframe} and open-source 3D models. The creation of this additional dataset was necessitated by the fact that the validation experiments were conducted in VR, and that this dataset provided a broader set of environments for evaluation.

The combined dataset contains 8 distinct object classes essential for navigation. The class IDs and instance counts are summarized in Table \ref{tab:class_ids}. Data augmentation (with horizontal image flipping) was applied to the synthetic frames of the additional dataset. 

\begin{table}[t]
\centering
\caption{Object Classes in the Combined Training Dataset.}
\label{tab:class_ids}
\begin{tabular}{|c|l|c|}
\hline
\textbf{Class ID} & \textbf{Class Name} &
\textbf{Instance Count}\\
\hline
0 & Person & 6126 \\
1 & Crossway & 18512 \\
2 & Potted Plant & 24872 \\
3 & Pole & 15382 \\
4 & Garbage & 4260 \\
5 & Bus Station & 5414 \\
6 & Car & 1231 \\
7 & Tree Trunk & 1550 \\
\hline
\end{tabular}
\end{table}

\begin{figure}[h!]
    \centering
    \begin{subfigure}[b]{1\linewidth}
        \centering
        \includegraphics[width=0.75\textwidth]{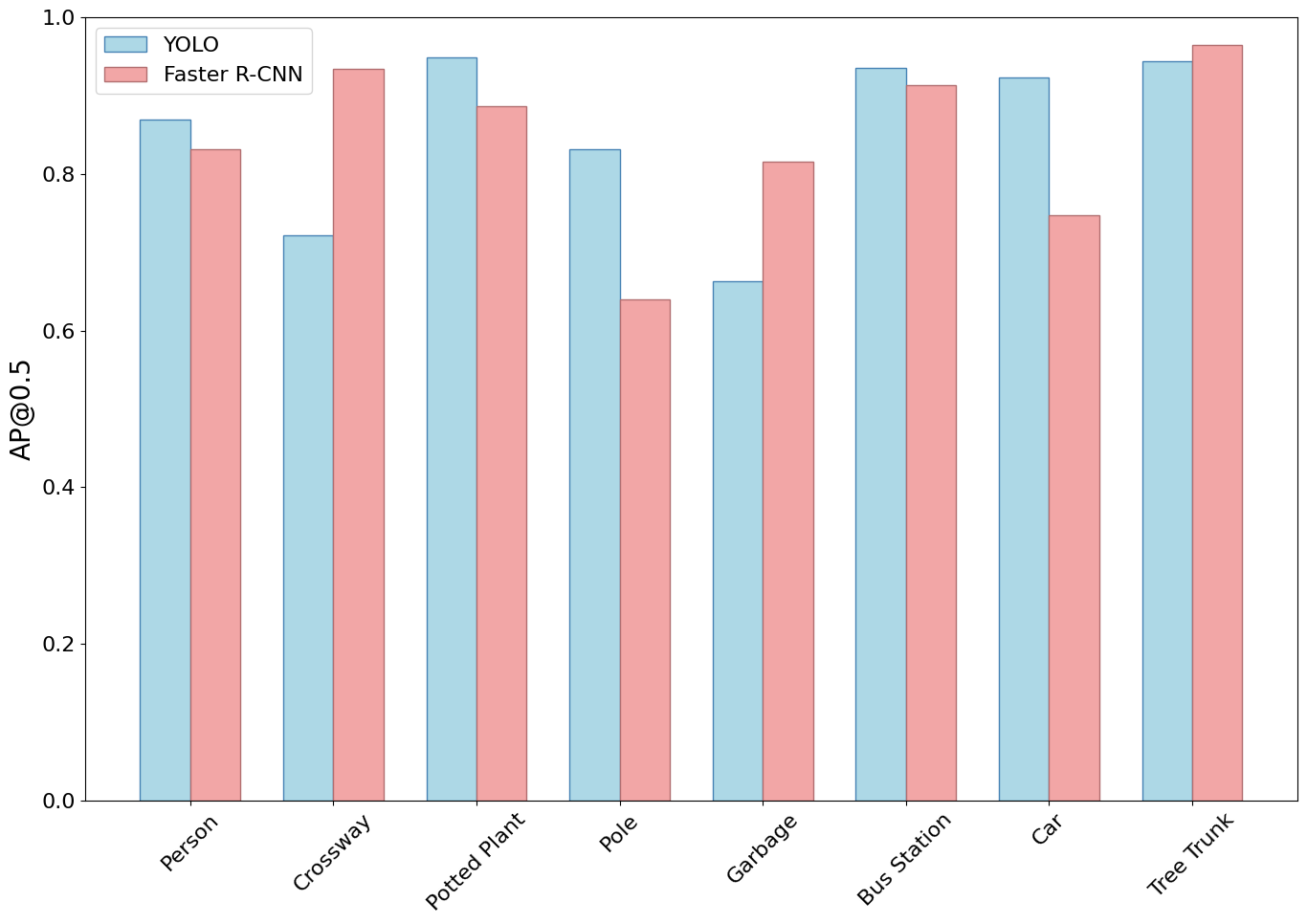}
        \caption{}
        \label{fig:map}
    \end{subfigure}
    
    \vspace{0.3cm} 
    
    \begin{subfigure}[b]{1\linewidth}
        \centering
        \includegraphics[width=0.75\textwidth]{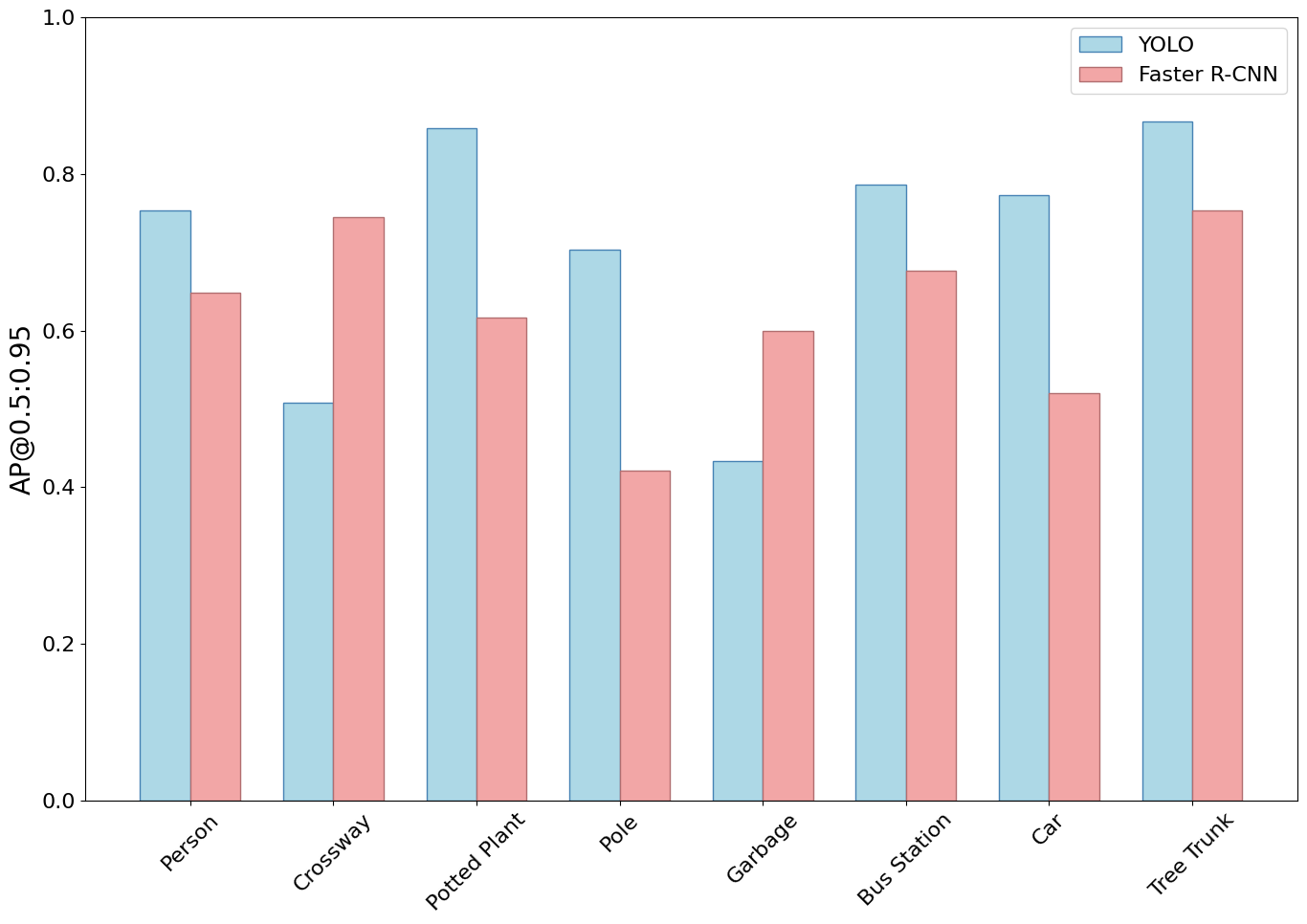}
        \caption{}
        \label{fig:coco}
    \end{subfigure}
    
    \caption{Mean average precision (mAP) for \textbf{YOLOv8s} and \textbf{Faster R‑CNN} detectors. (a) mAP@IoU=0.5 highlights the detection accuracy at the standard threshold, with both models shown side‑by‑side in distinct colors. (b) mAP@IoU=0.5:0.95 (COCO style) provides stricter evaluation across varying IoU levels.}
    \label{fig:model_comparison}
\end{figure}

For both detection models, the performance was evaluated using standard metrics, mainly mean average precision (mAP) at different intersection over union (IoU) thresholds and class-wise average precision at IoU=0.5 (AP@0.5). Fig. \ref{fig:model_comparison} shows the average precision across classes. The YOLOv8s model achieved a COCO-style mAP @ IoU[0.5:0.95] of 0.71, outperforming the Faster R-CNN model whose mAP was just 0.6224. Moreover, the YOLOv8s model had a lower inference time (120 ms per frame) compared to the R-CNN model (600 ms per frame).

The output of this pre-processing stage is a structured list containing bounding box coordinates, class IDs, and confidence scores for all detected objects per frame. This list serves as the primary input to the subsequent priority scoring and translation stage.



\subsubsection{Priority scoring and translation}
\label{subsec:translator}

The initial detection stage typically yields a high density of objects, often exceeding 10 instances per frame. Displaying this raw information on a prosthetic device with a limited FOV of $7^{\circ}$ and a limited number of electrodes of 378 \cite{PrimaSubretinal} results in visual saturation, as shown in Fig. \ref{fig:saturation}. To mitigate these limitations, three critical functions are performed in this stage:
\begin{figure}[h!]
    \centering
    \includegraphics[width=0.3\linewidth]{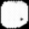}
    \caption{Illustration of phosphene saturation. Without prioritization, the display of raw detection data results in an unintelligible “white-out” effect due to the limited spatial resolution of the implant.}
    \label{fig:saturation}
\end{figure}

\begin{figure}[h]
    \centering
    \includegraphics[width=1\linewidth]{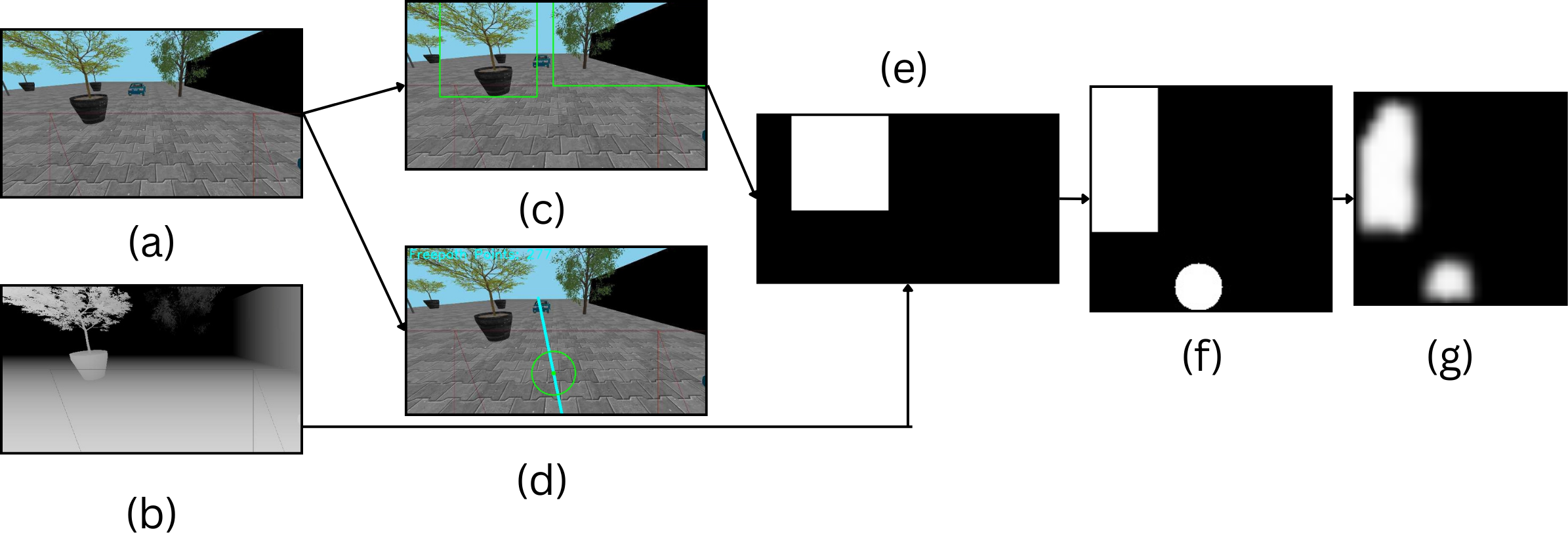}
    \caption{The visual translation and prioritization pipeline. This sequence demonstrates the “Safety Tier” filtering process with a maximum priority constraint of $k_{max}=1$ (Top-1 Threat). (a) Raw RGB input. (b) Corresponding depth map. (c) Semantic object detection (YOLOv8). (d) Free space extraction. (e) Priority masking \& abstraction: The system isolates the highest-scoring obstacle based on Eq. \ref{eq:scoring} and converts it into a high-contrast binary primitive. (f) FOV optimization \& path guidance: The map is cropped to a $30^{\circ}$ FOV and a visual cue (ball) is added to indicate the safe path. (g)Phosphene simulation: The final stimulation pattern sent to the implant.}
    \label{fig:scene_pipeline}
\end{figure}

\begin{enumerate}
    \item \textbf{Object prioritization:} 
    A priority-based filtering mechanism is proposed where each object is given a priority score based on its distance from the user. Then, only the top $k$ objects are selected, where $k$ is a hyperparameter defining the object density level (i.e., the maximum number of objects to be rendered for prosthetic vision). Two different values of $k$ (denoted in Section 3 as Level 1 and Level 2) were employed. The priority score $S(O_i)$ for an object $i$ is calculated as:
    
   \begin{equation}
        S(O_i) = \alpha \cdot I_{depth}(O_i)
        \label{eq:scoring}
    \end{equation}
    
    \noindent where $\alpha$ is a tunable scaling factor for intensity normalization, and  $I_{depth}(O_i)$ denotes the mean pixel intensity of object $i$ in the inverted depth map, normalized to $[0,1]$. In this representation, higher intensity values correspond to closer physical proximity.

    \item \textbf{Visual abstraction:} 
    Given the low spatial resolution of visual prosthetic devices, resolving complex object contours is perceptually difficult for the user, as demonstrated by the simulated prosthetic vision outputs of previous studies \cite{kucukoglu2025, BaylerExperiment, simpleNavigation}. Therefore, objects are abstracted into primitive shapes (e.g., bounding boxes) scaled by their relative physical size to maintain their aspect ratios. This simplifies the visual scene, providing the user with essential spatial cues—specifically location and presence—rather than high-frequency morphological details. This abstraction significantly reduces the mental workload required for navigation.

    \item \textbf{Field-of-view optimization \& path guidance:}
    Next, the abstracted map is cropped to the bottom-center quadrant of a $30^{\circ}$ FOV. This specific size was determined empirically, representing an optimal balance for collecting sufficient environmental data while preserving the patient's spatial perception. A wider FOV introduces excessive information that degrades spatial recognition, whereas a narrower FOV truncates critical object features. This $30^{\circ}$ digital buffer acts as a necessary intermediate step before mapping the scene to the PRIMA system's hardware constraints. Specifically, while the physical retinal implant is limited to a strict $7^{\circ}$ visual field, the system projects a $17^{\circ}$ FOV onto the retina to accommodate natural eye movements in the absence of active eye-tracking. The $30^{\circ}$ software crop ensures enough spatial context is retained before the final downsampling to this $17^{\circ}$ projection. Finally, a ``guiding ball'' is injected into the free space coordinates calculated in step (d) of Fig.~\ref{fig:scene_pipeline}, directing the user along the safe path.
\end{enumerate}

The complete visual transformation pipeline, illustrating the transition from raw input to the simplified safe output, is summarized in Fig. \ref{fig:scene_pipeline}.

\subsubsection{End-to-end stimulation parameter learning and phosphene map generation}
\label{subsec:pipeline2}


This stage transforms the simplified scene from the previous stage into optimized stimulation parameter values which, in reality, will be used to stimulate the retinal prosthesis for the blind to perceive the simplified scene. Since it is not feasible to stimulate the prosthetic vision device, a simulator is used to visualize what the blind patient will perceive if the implanted prosthetic device is stimulated by these values. 

The stimulation parameter used here is the amplitude. Hence, this stage aims to learn the correct amplitude array, which has the same length as the PRIMA's number of electrodes.

Since this is an optimization problem, the autoencoder architecture is used as proposed by Granley et al \cite{granley2022}.

\begin{figure}[h!]
    \centering
    \includegraphics[width=1\linewidth]{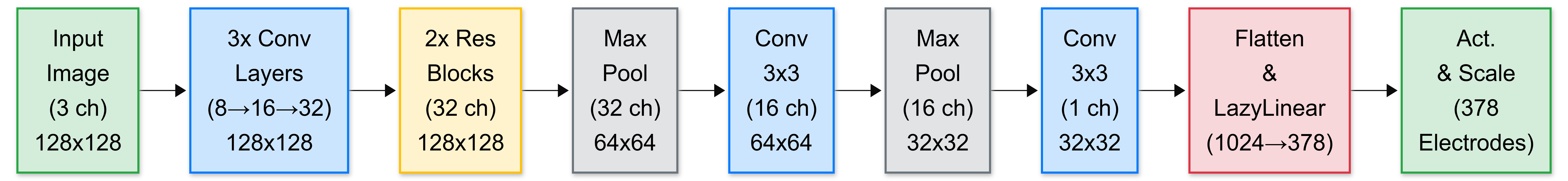}
    \caption{The CNN Encoder Architecture.}
    \label{fig:EncoderArchitecture}
\end{figure}

\paragraph{Encoder}
The simplified scene is fed into a convolutional neural network (CNN) encoder. The architecture used is inspired by the architecture proposed by K\"u\c{c}\"uko\u{g}lu et al. \cite{kucukoglu2025}, but it was simplified to have only two residual blocks and a reduced number of convolution layers, as shown in Fig. \ref{fig:EncoderArchitecture}. Its primary function is to derive amplitude values from the input features; these amplitudes serve as the electrical stimulation values that will be fed into the electrodes of the retinal prosthetic device.

\paragraph{Simulator}
The simulator (biological decoder) is a critical modeling component; it outputs the perceptual response (phosphene map) from the electrical stimulation values. This allows system validation without the risks associated with surgical trials. The simulator takes the amplitude array obtained from the encoder to generate the phosphene map. 

 The \textit{pulse2percept} library was utilized, which is an open-source library designed for visual prosthetic simulation \cite{Beyeler2017}. Since PRIMA is a subretinal implant \cite{PrimaSubretinal}, the resulting phosphene perception is described as continuous white lines\cite{park2025simulation}, and patients who suffered age-related macular degeneration (AMD) when had the implant at the central location within the macular region experienced visual acuity very close to the pixel size proposed \cite{palanker2022simultaneous}. To replicate this perception, the \textit{scoreboard spatial perceptual model} was used, providing a reliable spatial approximation of the phosphene geometry induced by photoreceptor stimulation. This perceptual model was re-implemented in PyTorch to have a fully differentiable training loop.

\begin{figure}[h!]
    \centering
    \begin{subfigure}[b]{0.23\linewidth}
        \centering
        \includegraphics[height=1.7cm, width=\linewidth, keepaspectratio]{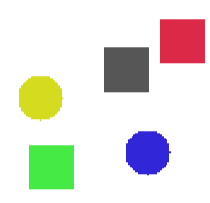}
        \caption{}
        \label{fig:coloredSOC}
    \end{subfigure}\hfill
    \begin{subfigure}[b]{0.23\linewidth}
        \centering
        \includegraphics[height=1.7cm, width=\linewidth, keepaspectratio]{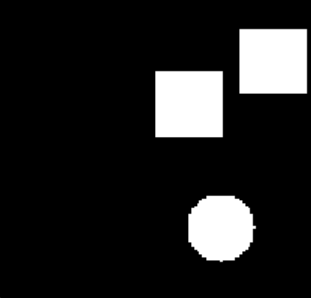}
        \caption{}
        \label{fig:BinarizedSOC}
    \end{subfigure}\hfill
    \begin{subfigure}[b]{0.23\linewidth}
        \centering
        \includegraphics[height=1.7cm, width=\linewidth, keepaspectratio]{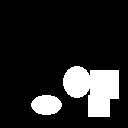}
        \caption{}
        \label{fig:CustomizedSimpleData}
    \end{subfigure}\hfill
    \begin{subfigure}[b]{0.23\linewidth}
        \centering
        \includegraphics[height=1.7cm, width=\linewidth, keepaspectratio]{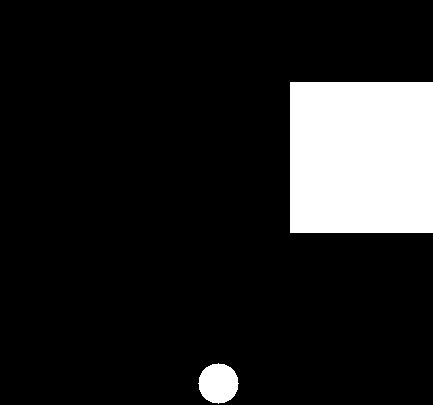}
        \caption{}
        \label{fig:realSimpleScene}
    \end{subfigure}
    
    \caption{Data preprocessing: (a) a colored Sort-of-CLEVR image, (b) the preprocessed Sort-of-CLEVR image, (c) a synthetic image, (d) the target simplified scene.}
    \label{fig:data_progression}
\end{figure}

\paragraph{Pipeline Training}
To optimize the CNN encoder for accurate amplitude prediction, the model was trained on a hybrid dataset designed to approximate the properties of the proposed simplified scenes. The training data consisted of the Sort-of-CLEVR dataset and a customized dataset. The Sort-of-CLEVR dataset contains simple random geometric primitives (squares and circles), originally developed for relational reasoning tasks \cite{sortofclever}. To align data samples from this dataset with the target input data, specific preprocessing steps were applied, including binarization and color inversion. As illustrated in Fig.\ref{fig:coloredSOC}, the original CLEVR image is shown, whereas Fig.\ref{fig:BinarizedSOC} shows the corresponding preprocessed image. The customized dataset was generated to bridge the domain gap between standard geometric shapes and specific primitive shapes used in the navigation task. This dataset was created by randomly generating rectangles and ellipses with varying aspect ratios and degrees of overlap as in Fig.\ref{fig:CustomizedSimpleData}. These shapes were designed to mimic the simplified representations of obstacles as shown in Fig. \ref{fig:realSimpleScene}.

The core objective of this pipeline is to minimize the perceptual difference between the input semantic map (simplified scene) and the simulated phosphene output. To achieve this objective, a compound loss is defined as follows:
\begin{equation}
    \mathcal{L}_{\mathrm{total}} =
    \lambda_{\mathrm{MSE}}\mathcal{L}_{\mathrm{MSE}}(P, I)
    + \lambda_{\mathrm{Dice}}\mathcal{L}_{\mathrm{Dice}}(P, I)
    + \lambda_{\mathrm{VGG}}\mathcal{L}_{\mathrm{VGG}}(P, I)
    + \lambda_{\mathrm{Smooth}}\mathcal{L}_{\mathrm{Smooth}}(P)
\end{equation}

where $I$ represents the set of input semantic maps (ground-truth target images), and $P$ represents the set of corresponding generated phosphene maps (percepts). The parameters $\lambda_{\mathrm{MSE}}$, $\lambda_{\mathrm{Dice}}$, $\lambda_{\mathrm{VGG}}$, and $\lambda_{\mathrm{Smooth}}$ are weighting coefficients that control the relative contribution of each loss component to the total objective. The four loss components are defined as follows:

\begin{itemize}
    \item \textbf{Mean Squared Error (MSE):} this loss ensures the pixel-wise fidelity and is defined as:
    \begin{equation}
        \mathcal{L}_{MSE}(P,I) = \frac{1}{N} \sum_{i=1}^{N} (P_i - I_i)^2
    \end{equation}
    where $N$ is the number of training samples.

    \item \textbf{Dice Loss:} Each percept image has a black background overlaid with white percepts, resembling conventional image segmentation masks. Hence, the Dice loss, widely used in segmentation tasks, forces the encoder to care about the white pixels (the shape) much more than the black background. This loss can be mathematically defined as:
    \begin{equation}
        \mathcal{L}_{Dice}(P,I) = 1 - \frac{2 \sum (P \cdot I) + \epsilon}{\sum P + \sum I + \epsilon}
    \end{equation}
    where $\epsilon$ is a small factor to prevent division by zero.

    \item \textbf{Perceptual Loss (VGG):} This loss measures high-level semantic similarity using features extracted from a pre-trained VGG19 network, following the approach introduced by Granley et al \cite{granley2022}. The loss is defined as:
    \begin{equation}
        \mathcal{L}_{VGG} = \beta \cdot \| \phi(P) - \phi(I) \|_2^2
    \end{equation}
    where the operator $\phi$ extracts VGG features up to the layer \texttt{relu3\_3} and $\beta = 0.1$ scales the loss.

    \item \textbf{Smoothness Loss:} his loss regularizes the output and reduces noise by penalizing the second spatial derivative (Laplacian). This loss is defined as:
    \begin{equation}
        \mathcal{L}_{Smooth} = \frac{1}{N} \| \nabla^2 P \|_2^2
    \end{equation}
    where $\nabla^2$ denotes the convolution operator with a Laplacian kernel.
\end{itemize}

\begin{figure}[h!]
    \centering
    \includegraphics[width=\linewidth]{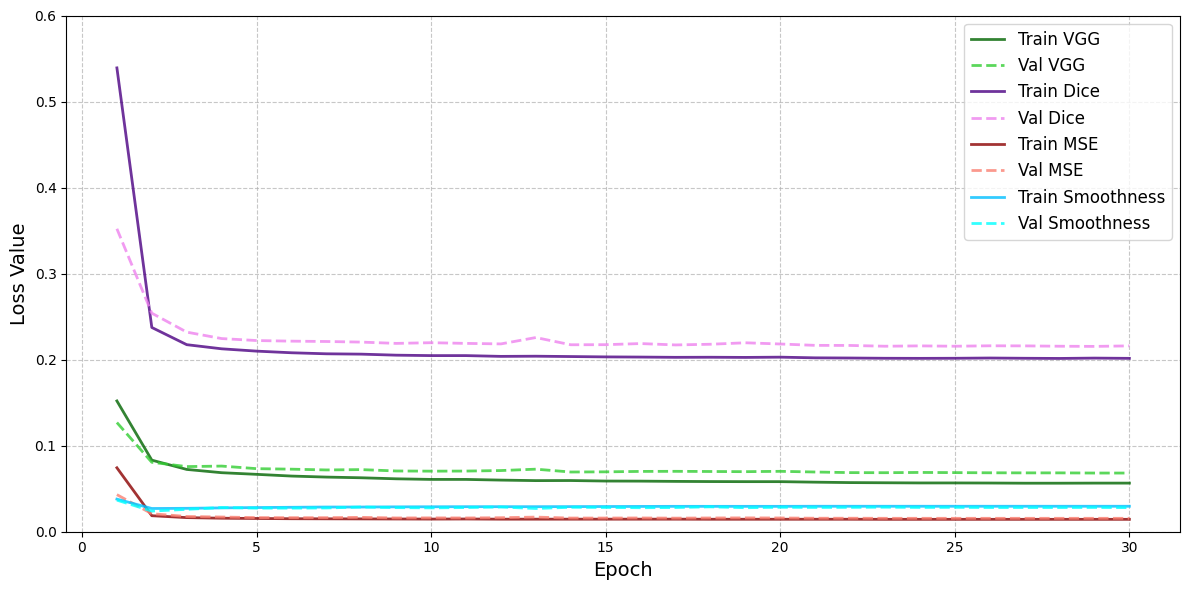}
    \caption{Encoder training and validation results using the compound loss, where losses are weighted using weight normalization.}
    \label{fig:training_res}
\end{figure}

After experimenting with different combinations of losses and weights, it was found that weighting the loss terms using weight normalization yields the best validation results, as shown in Fig \ref{fig:training_res}.

\subsection{Dawn Pilot: The VR Research Platform}
\label{subsec:architecture}

To safely validate the generated phosphene maps without the risks associated with human surgical trials, a custom, web-based virtual reality platform named ``Dawn Pilot" is proposed. Designed specifically as a clinical research tool, this platform addresses the need for safe, reproducible, and highly measurable simulated prosthetic vision (SPV) experiments.

\begin{figure}[htbp]
    \centering
    \begin{subfigure}[b]{0.48\linewidth} 
        \centering
        \includegraphics[width=\linewidth]{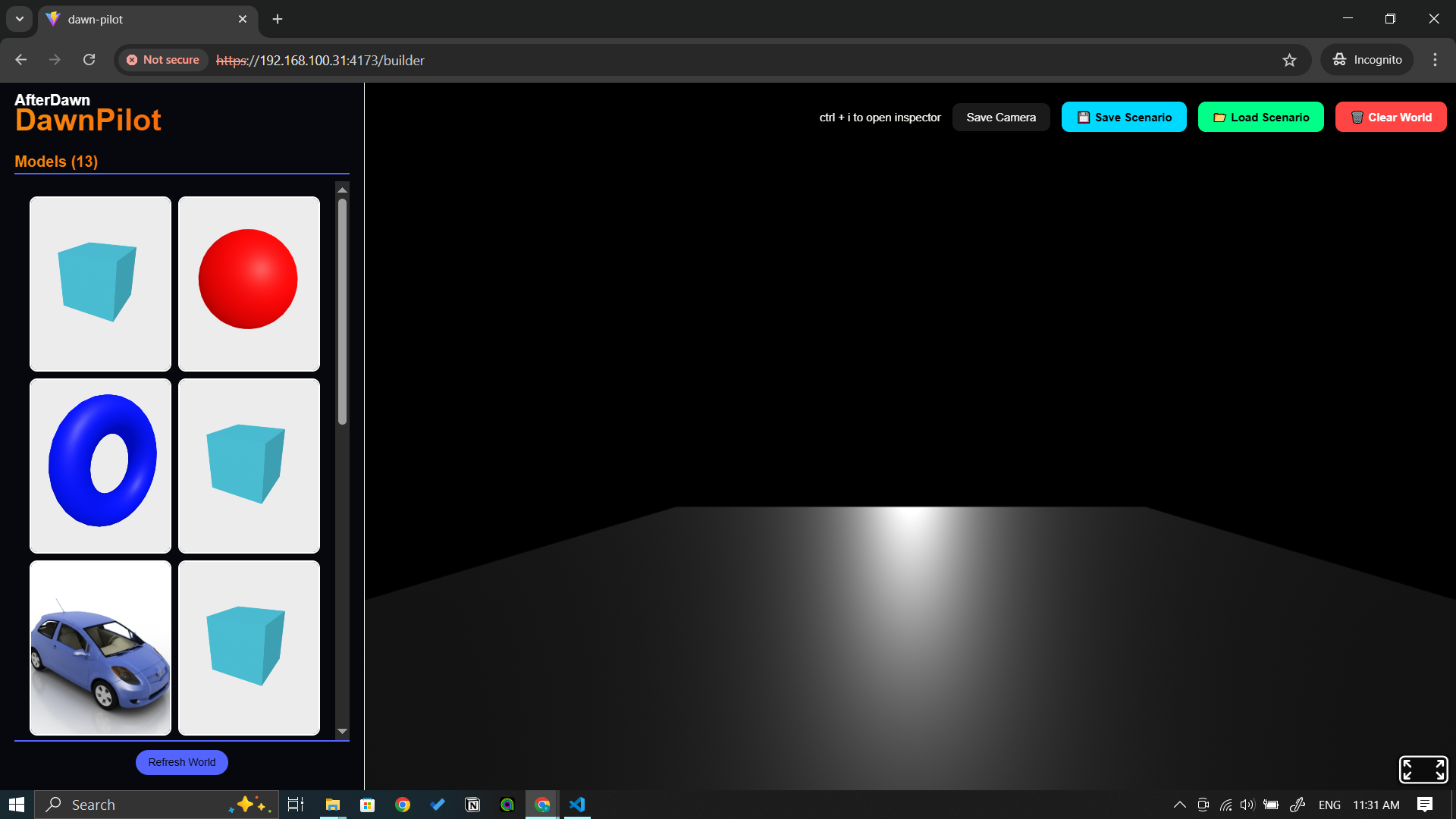}
        \caption{Initial Environment}
        \label{fig:builder_a}
    \end{subfigure}
    \hfill 
    \begin{subfigure}[b]{0.48\linewidth}
        \centering
        \includegraphics[width=\linewidth]{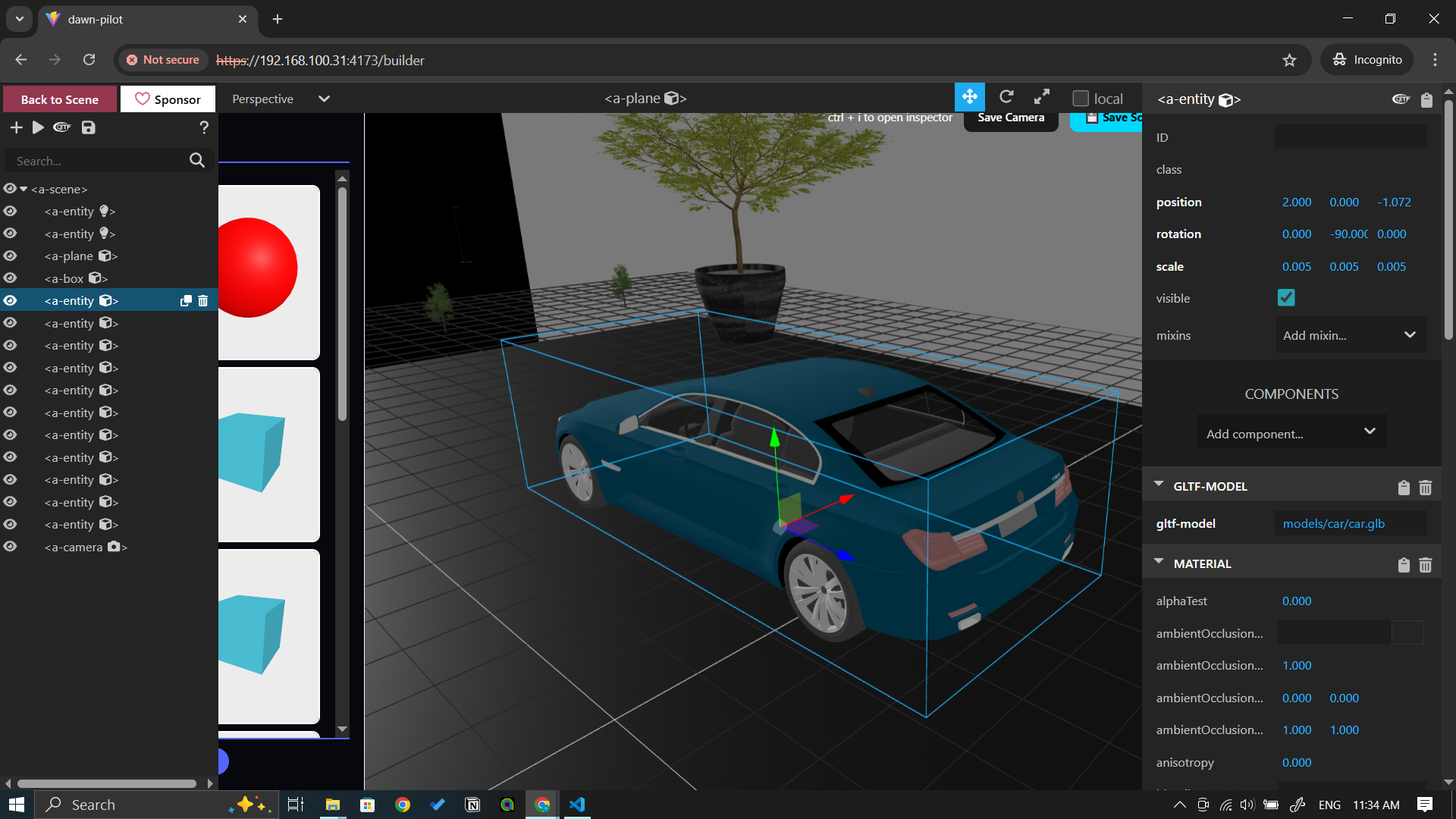}
        \caption{Entity Configuration}
        \label{fig:builder_b}
    \end{subfigure}
    
    \caption{The custom Scenario Builder workflow for experimental reproducibility. \textbf{(a)} A clean environment is initialized to build experiment scenarios. \textbf{(b)} Different entities (e.g., obstacles, waypoints) are added, and their 3D spatial coordinates are manually configured before the scenario is saved to the Experiment Vault for VR client synchronization.}
    \label{fig:scenario_builder}
\end{figure}

To ensure environmental reproducibility, the platform features a ``Scenario Builder" graphical user interface (GUI) (Fig. \ref{fig:scenario_builder}). This GUI allows researchers to construct standardized 3D testing environments—such as streets populated with obstacles (e.g., trash bins, trees, and parked cars). Entities can be added via a simple insertion menu by specifying their initial spatial coordinates, with the added flexibility of dragging and repositioning any object post-placement. These configurations are then saved to a JSON file containing all the data required to recreate the environment, ensuring the exact same spatial layout can be reliably loaded for multiple subjects.

To facilitate human-in-the-loop testing, the platform provides a ``Researcher Control" dashboard, which allows the investigator to monitor the subject's live VR view and the AI's processed output in real time. Furthermore, the architecture is equipped with an automated telemetry system designed to eliminate manual observation errors during any VR session. This system continuously logs spatial and temporal metrics, including path deviation (recorded via continuous camera position tracking relative to the optimal spatial trajectory), collision events (detected via A-Frame’s bounding boxes whenever the user’s virtual body intersects with a static obstacle), and session time (the total time taken by a subject to complete a specific navigational course designed by the researcher).

\subsubsection{\textbf{Distributed Architecture and Latency Mitigation}}
While immersive VR is excellent for validation, executing real-time object classification, priority scoring, semantic translation, and phosphene mapping simultaneously poses a massive computational burden. If processed locally on a single machine, this heavy pipeline introduces severe rendering latencies. In VR, visual latency disconnects the user’s vestibular system from their visual feedback, which can induce motion sickness and invalidate the experimental data \cite{UGUR2022768}. 

\begin{figure}[htbp]
    \centering
    \includegraphics[width=1\linewidth]{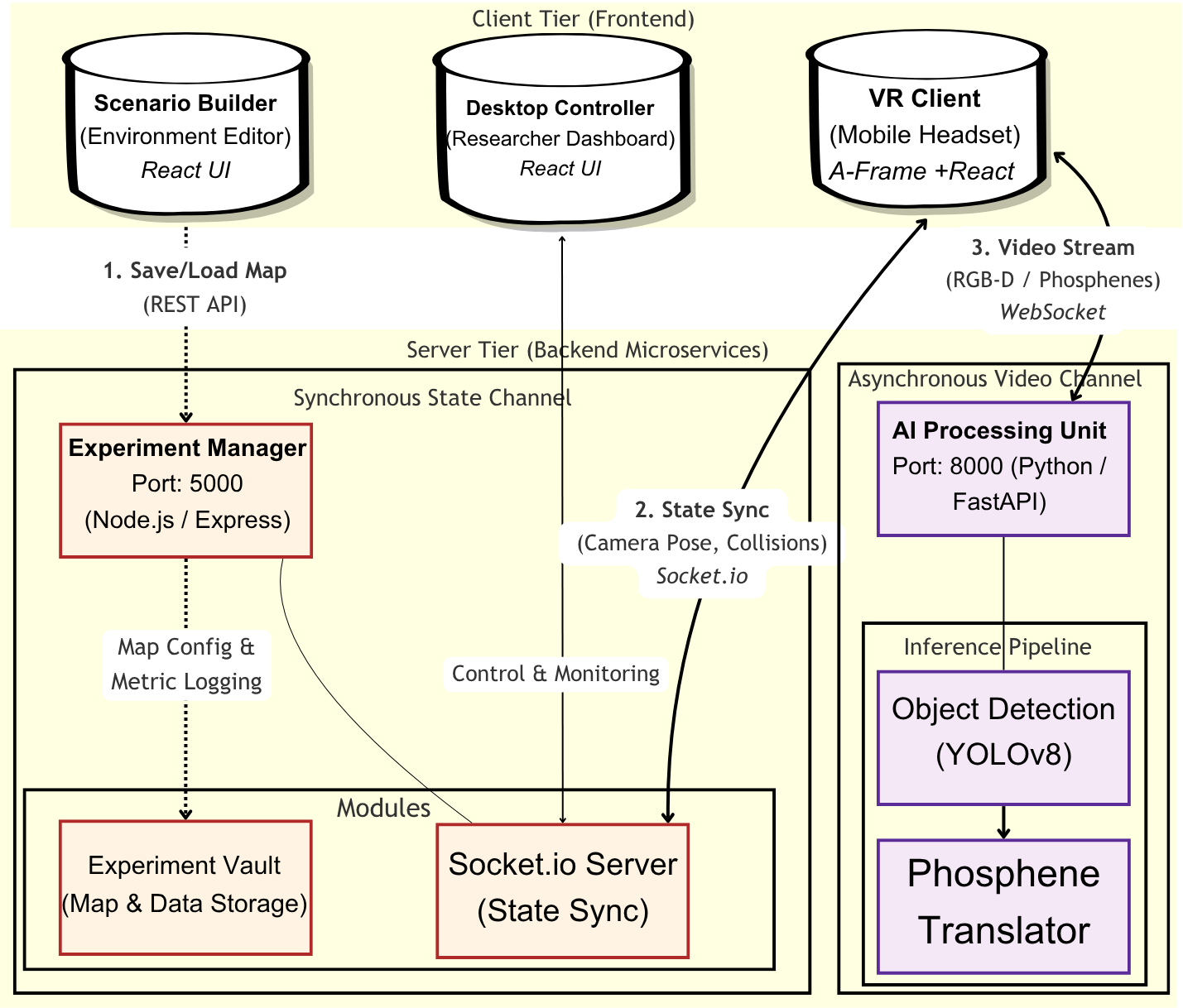}
    \caption{System Architecture. The framework utilizes a split-stack design to ensure low latency during immersive trials. The Client Tier introduces a \textit{Scenario Builder} for reproducible environment generation. The Server Tier decouples synchronous state management and telemetry logging (Experiment Manager) from heavy asynchronous deep learning inference (AI Processing Unit).}
    \label{fig:architecture}
\end{figure}

To maintain the strict latency constraints required for comfortable SPV, a distributed, three-tier microservice architecture is engineered as shown in Fig.~\ref{fig:architecture}. Specifically, the Client Tier (VR Frontend) is built on \textit{A-Frame (WebXR)} \cite{aframe} and \textit{React}, handling only stereoscopic rendering and head tracking. By keeping the client lightweight, a stable 30-FPS visual loop was maintained to minimize motion sickness within hardware constraints. 

To coordinate the operation of this frontend with the rest of the system, the Synchronization (Server) Tier employs a central \textit{Node.js/Express} server utilizing \textit{Socket.io} to maintain a consistent ``World State.'' This tier efficiently manages lightweight, event-driven data, such as tracking camera coordinates, logging collisions, and syncing with the Researcher Dashboard. 

Finally, all heavy computational workload is offloaded to the Inference Tier, which operates as a dedicated \textit{FastAPI (Python)} microservice. This service hosts the YOLOv8n object detector alongside the phosphene translation algorithms. This tier processes the incoming frame stream asynchronously image-by-image, applies the optimized stimulation values, and then returns the generated phosphene map back to the client.

To further minimize the end-to-end latency, a dual-channel communication strategy was implemented. A dedicated, high-throughput WebSocket stream transmits encoded RGB-D frames with corresponding z-buffer depth maps directly to the Inference Tier, bypassing the Synchronization Tier entirely. This separation prevents heavy video traffic from bottlenecking the state management logic, ensuring real-time, low-latency prosthetic vision for the user.




\section{Experimental Environment, and Setup}
Utilizing the Dawn Pilot VR platform described in Section \ref{subsec:architecture}, we designed a dynamic navigational task to evaluate the efficacy of the proposed semantic translation and prioritization algorithms with sighted human subjects.

\begin{figure}[htbp]
    \centering
    \begin{subfigure}[b]{0.45\linewidth}
        \centering
        \includegraphics[width=0.48\linewidth]{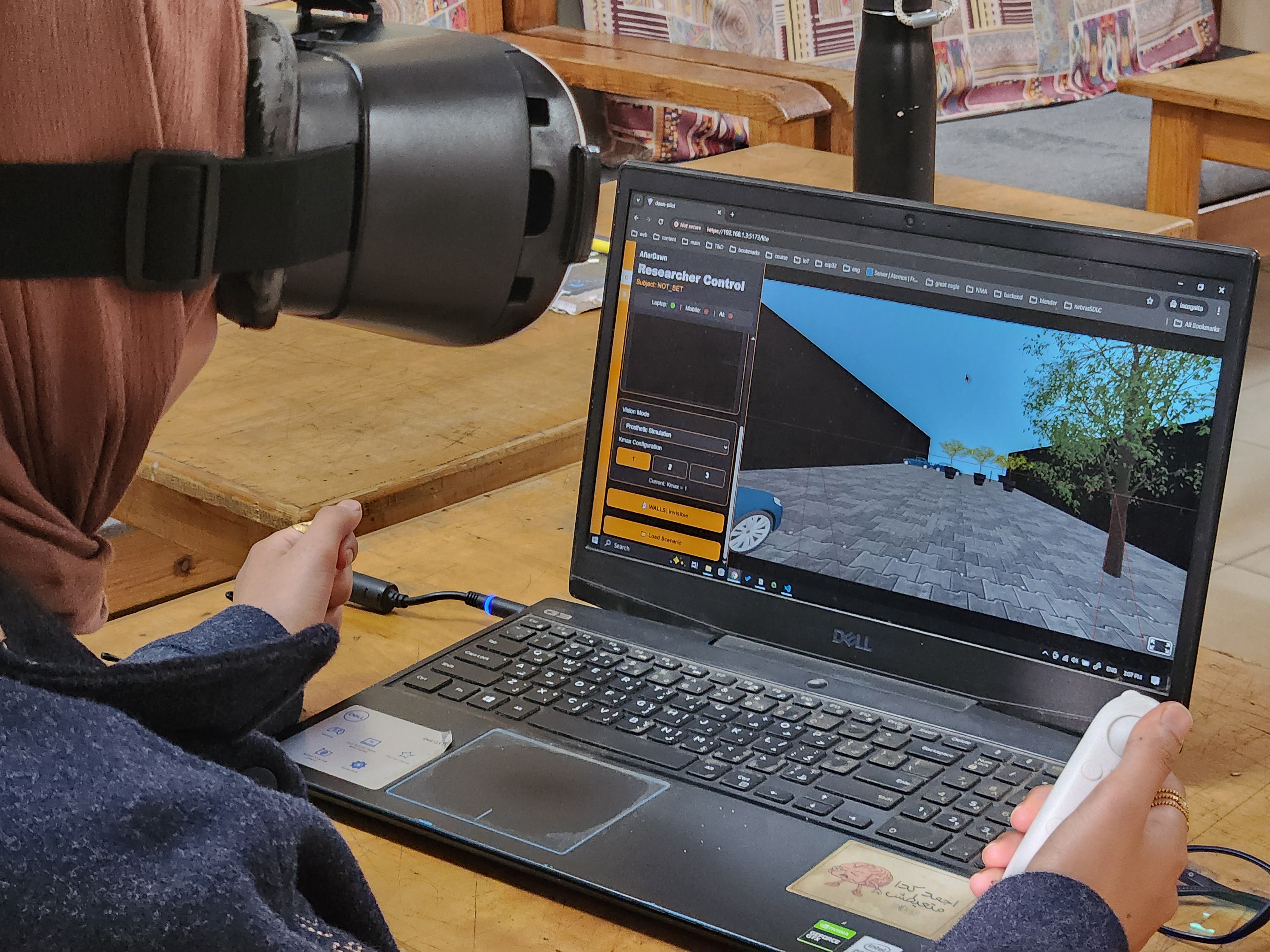} \hfill
        \includegraphics[width=0.48\linewidth]{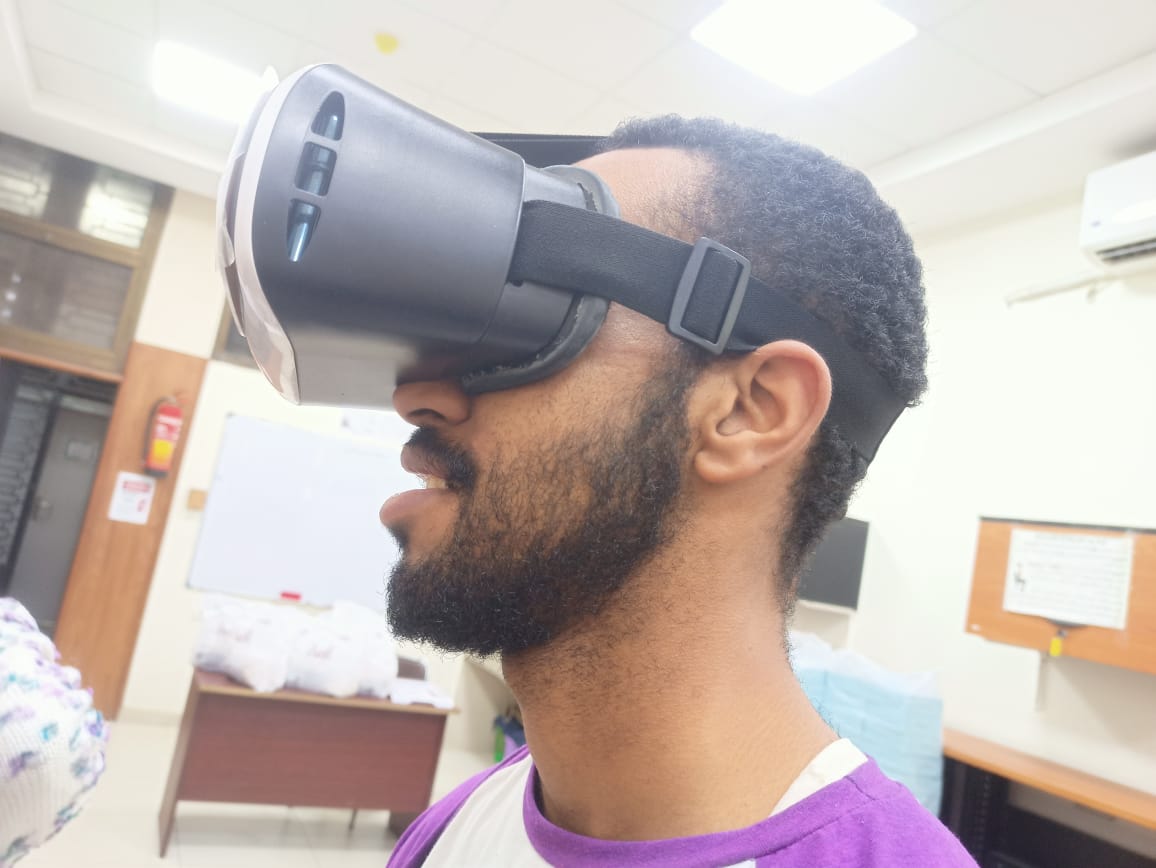} \\ 
        \vspace{1mm} 
        \includegraphics[width=0.48\linewidth]{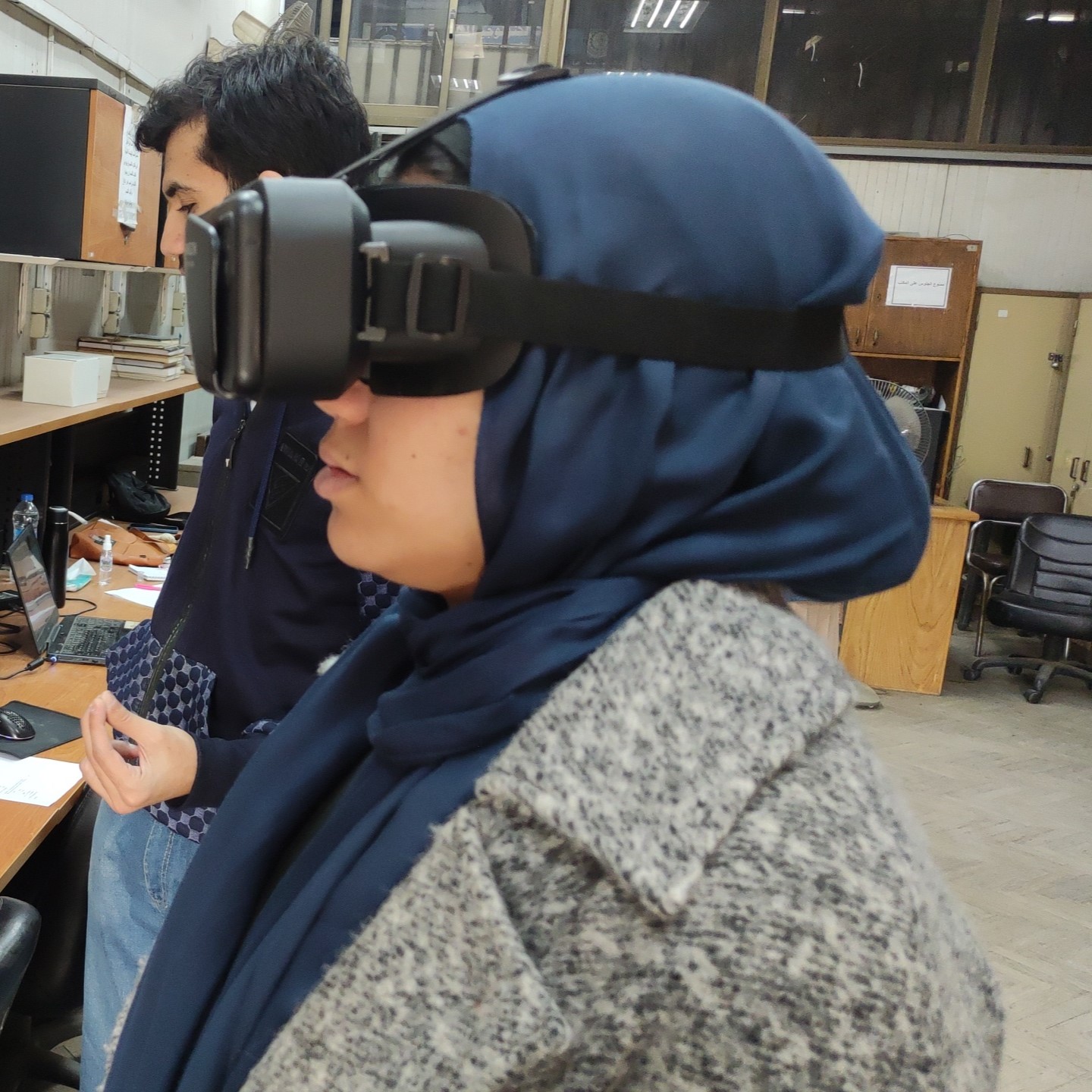} \hfill
        \includegraphics[width=0.48\linewidth]{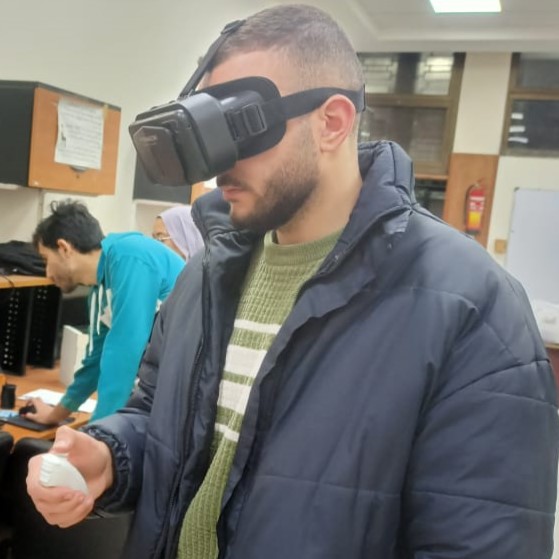}
        \caption{}
        \label{fig:participants_grid}
    \end{subfigure}
    \hfill 
    \begin{subfigure}[b]{0.53\linewidth}
        \centering
        \includegraphics[width=\linewidth]{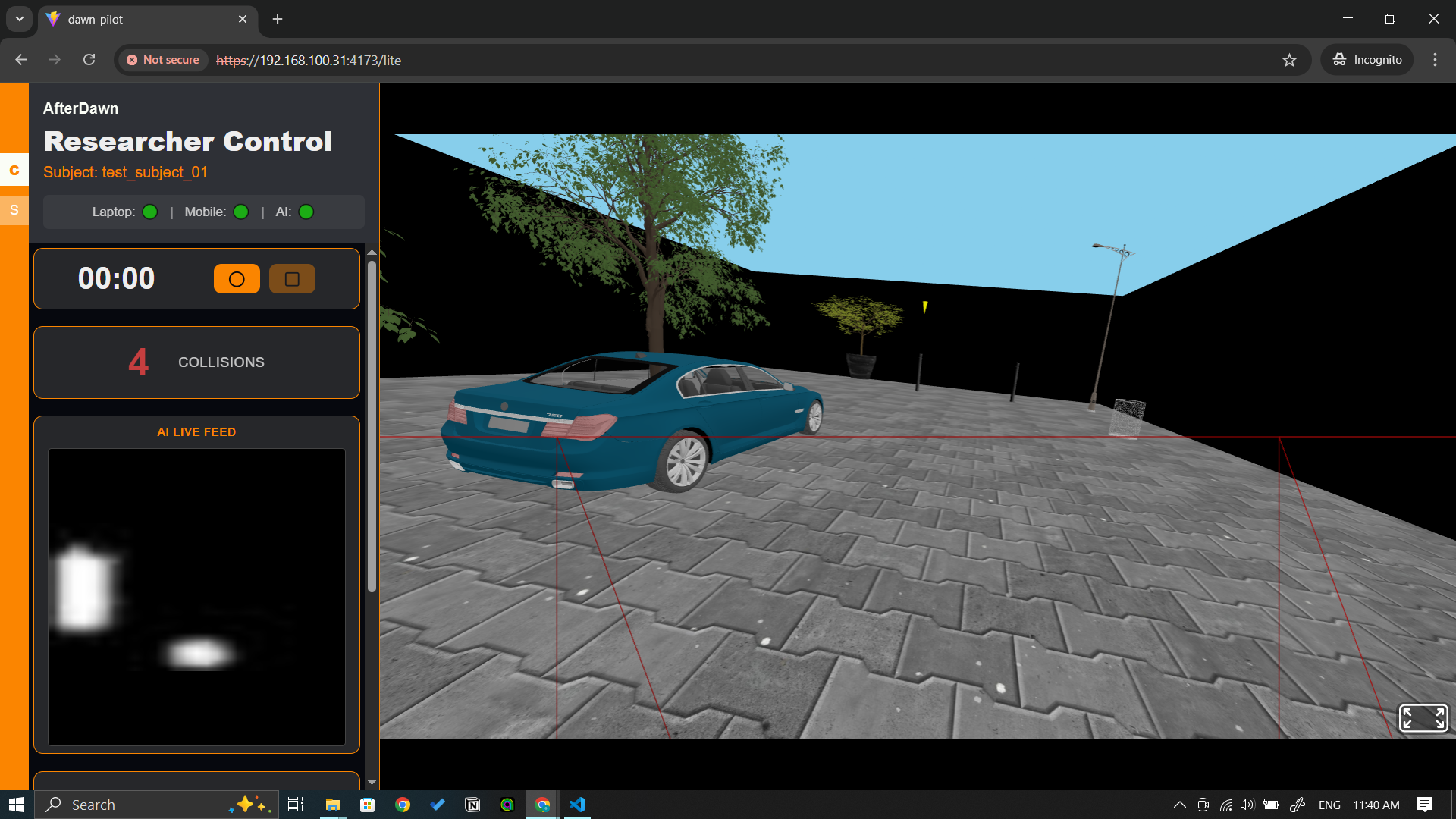}
        \caption{}
        \label{fig:vr_view}
    \end{subfigure}
    
    \caption{Experimental Validation. \textbf{(a)} A subset of the 14 participants performing the navigation task using the Bluetooth controller. \textbf{(b)} The Researcher Dashboard, displaying the live VR environment alongside its corresponding generated phosphene map to track subject interaction, alongside session controls.}
    \label{fig:experiment_composite}
\end{figure}

\subsection{Participants}
A total of 12 volunteers participated in our experiment. All participants possessed normal or corrected-to-normal vision and had no prior experience with phosphene simulation. The participants consisted of 6 males and 6 females, with an age range of 14 to 26 years. None of the participants had a history of epilepsy, seizures, fainting spells, motion sickness, or severe reactions to flashing lights.

\subsection{Virtual environment and task objectives}
\begin{figure}[htbp]
    \centering
    \begin{subfigure}[b]{0.49\linewidth}
        \centering
        \includegraphics[height=2.5cm, width=\linewidth, keepaspectratio]{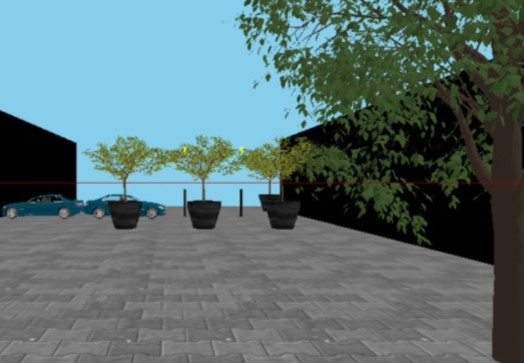}
        \caption{}
        \label{fig:obstacles}
    \end{subfigure}
    \hfill
    \begin{subfigure}[b]{0.49\linewidth}
        \centering
        \includegraphics[height=2.5cm, width=\linewidth, keepaspectratio]{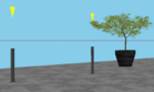}
        \caption{}
        \label{fig:target}
    \end{subfigure}
    \caption{Scene examples highlighting key navigational elements: \textbf{(a)} Static obstacles, and \textbf{(b)} The target destination (the poles with yellow arrow on top).}
    \label{fig:scenes}
\end{figure}

The simulated environment was constructed to depict urban settings with streets and static obstacles (including cars, trees, poles and potted plants) as shown in Fig. \ref{fig:obstacles}. To prevent participants from relying on path memorization, three distinct scene layouts were generated. The target destination for each scene was highlighted by floating arrows as shown in Fig. \ref{fig:target}. For every trial, the primary objective was to navigate the environment and reach the destination as fast as possible while minimizing the number of collisions with obstacles.

To accomplish this task, the simulation was displayed on a mobile phone inserted into a VR headset, enabling participants to navigate through the scenes using a hybrid control scheme. Specifically, this scheme paired a Bluetooth controller for translation (providing 2 degrees of freedom (DoF) for forward, backward, left, right, and diagonal movement) with natural head tracking (providing three additional DoF to allow users to look around the environment in all directions).

\subsection{Visual conditions}
\begin{figure}[htbp]
    \centering
    \begin{subfigure}[b]{0.32\linewidth}
        \centering
        \includegraphics[height=2cm, width=\linewidth, keepaspectratio]{figures/scene2.jpg}
        \caption{}
        \label{fig:normal}
    \end{subfigure}\hfill 
    \begin{subfigure}[b]{0.32\linewidth}
        \centering
        \includegraphics[height=2cm, width=\linewidth, keepaspectratio]{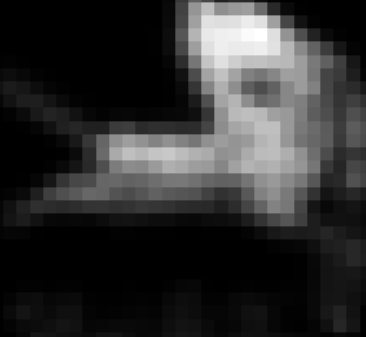}
        \caption{}
        \label{fig:edge}
    \end{subfigure}\hfill 
    \begin{subfigure}[b]{0.32\linewidth}
        \centering
        \includegraphics[height=2cm, width=\linewidth, keepaspectratio]{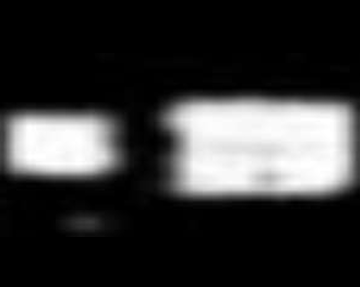}
        \caption{}
        \label{fig:phosphenes}
    \end{subfigure}
    
    \caption{Visual modes for simulated prosthetic vision (SPV): (a) Normal vision, (b)conventional edge detection solution, and (c) proposed smart navigation solution.}
    \label{fig:modes}
\end{figure}

Participants performed the navigation task under three different visual modes (conditions) (Fig. \ref{fig:modes}), which were designed to evaluate the effectiveness of our proposed system against standard processing techniques. 

The Normal Vision mode presented a standard 3D scene restricted to a $17^\circ$ field of view (FOV). This mode served as a baseline to measure how quickly participants could adapt to the limited FOV inherent in the PRIMA implant \cite{PrimaSubretinal}. 

This baseline was thus compared against a conventional prosthetic simulation mode based on edge detection. Adapted from the solutions proposed by Kasowski et al. \cite{kasowski2025}, this mode fed raw scenes into a Canny edge detector to extract structural outlines. The resulting map was then projected directly to the encoder without any object simplification or prioritization. 

Finally, participants navigated using the proposed Smart Navigation mode, where scenes were processed through our novel pipeline. In contrast to the raw edge outlines of the edge detection mode, this mode utilized object detection and depth analysis to prioritize only the most relevant obstacles and safe paths before converting them into abstracted, simplified phosphene patterns.

\subsection{Experimental protocol}
The experiments always commenced with the Normal Vision mode. This initial phase served as our baseline to gauge subject performance and acclimatization to the $17^\circ$ FOV. Subsequently, the Edge Detection and Smart Navigation trials were randomized to avoid order bias and learning effect. To further prevent path memorization, the three distinct scene layouts were randomly distributed across these visual conditions for every participant. 

For each subject, eight navigation trials were carried out and recorded. Specifically, participants completed two trials using the Normal Vision baseline and two trials using the Edge Detection mode. Finally, under the Smart Navigation mode, participants completed four trials (with two trials for each of two different obstacle density levels).

The obstacle density is the maximum number of objects the system will show to the subject. The two density levels were: 1 object (level 1) and 3 objects (level 2). The duplication of each trial ensured better data reliability across the testing cohort.

\subsection{Evaluation and subject assessment}
As outlined in Section \ref{subsec:architecture}, the Dawn Pilot platform automatically recorded quantitative metrics for every trial to assess system efficiency, specifically tracking completion time (sec), number of collisions, and path deviation.

After all physical trials were completed, participants filled out an adapted NASA Task Load Index (NASA-TLX) questionnaire \cite{nasa_tlx_web} to assess cognitive load. To simplify the rating process, the scale was reduced to a 3-point system. Each participant rated the task across six dimensions: mental demand, physical demand, temporal demand, overall effort, performance satisfaction, and frustration level. Additionally, qualitative feedback was gathered regarding the perceived pros and cons of both the smart navigation and edge detection modes.

\section{Results}

For statistical analysis of the results, the different trials resulting from tuning the k parameter of the proposed Smart Navigation methodology were treated as two separate conditions rather than as a nested factor.  The analysis was divided into quantitative and qualitative components. For quantitative analysis, distinct statistical tests were used to answer specific research questions. In the qualitative analysis, participants’ survey notes were examined using thematic analysis.

\subsection{Comparing conventional and proposed solutions using the Friedman test}

First, to determine whether to average the two trials associated with each mode or use only the second trial, a Wilcoxon signed-rank test was conducted to compare collision counts and completion times of the two trials (denoted as Trial 1 and Trial 2). The analysis revealed a statistically significant reduction in completion time only in the second trial (p = 0.0192). Therefore, the Friedman test was conducted using Trial 2 data only to minimize the influence of early familiarization effects and to better represent stabilized user performance.

A Friedman test was used to determine whether there were differences in the number of collisions or completion time across navigation modes (conventional mode, proposed mode Level 1 (Smart Level 1), and proposed mode Level 2 (Smart Level 2)). The results indicated a statistically significant effect of the navigation  mode on collision rates ($\chi^2$(2) = 6.47, p = 0.039, Kendall’s W = 0.27).

Post-hoc pairwise comparisons using Wilcoxon signed-rank tests with Holm–Bonferroni correction revealed a significant difference between Smart Level 1 and Smart Level 2 (p = 0.033), with Smart Level 2 resulting in fewer collisions. However, no statistically significant differences were observed between the conventional baseline and Smart Level 1 (p = 0.165) or Smart Level 2 (p = 0.682).

Furthermore, no statistically significant differences were found among completion times across navigation modes.

\subsection{Generalized linear mixed model (GLMM) analysis for multi-trial performance evaluation }
While the initial tests provided a non-parametric comparison of trial-specific performance, a generalized linear mixed model (GLMM) framework was implemented by integrating all experimental trials into a single model. This approach accounted for the hierarchical structure of the data and the variability between participants, allowing for the quantification of fixed effects (navigational modes)  and random effects (inter-subject variability).
The primary objective of the GLMM analysis was to model the effect of the proposed levels (K1 and K3) and the conventional mode relative to the baseline Normal Vision mode. This analysis was used for evaluating the rate of change across the system, as it treats the trial index as a continuous predictor, thereby isolating the learning effect from the core functional efficacy of the navigational modes. The analysis provides an estimation of the incident rate ratio (IRR), quantifying the factor of increase or decrease in navigational risk and efficiency across all operational conditions.

\subsubsection{Task success}
The success metric was not very informative, since only one trial out of 95 failed (98.9\% success). With no enough variation, this metric cannot thus serve as a discriminative measure.

\subsubsection{Navigational safety: Collision analysis}


The descriptive statistics in Table \ref{tab:collision_analysis} show that the conventional solution was the least-collision option in the trials with a mean of 1.96 $\pm$ 3.54 collisions per trial and 14 collision-free trials. Level 2 of the proposed solution averaged 3.60 $\pm$ 4.71 collisions (median 1.5 collisions), with only four collision-free trials. Level 1 had a higher mean (4.39 $\pm$ 7.10 collisions) but showed better peak performance with nine collision-free trials. The high mean in Level 1 is primarily due to an outlier trial with 35 collisions, which skewed the data. The baseline `Normal Vision' mode was significantly safer, with 16 collision-free trials.
Since the collisions are a countable discrete metric and given the overdispersion of the data, a negative binomial GLMM model with a log-link was utilized to analyze the data in Python and R . The following formula was used:
\begin{equation}
\text{Collision count} \sim \mathrm{Mode} + \mathrm{Trial\ index} + \mathrm{Scenario} + (1 \mid \mathrm{Subject})
\end{equation}

\begin{figure}[!ht]
    \centering
    \includegraphics[width=.5\linewidth]{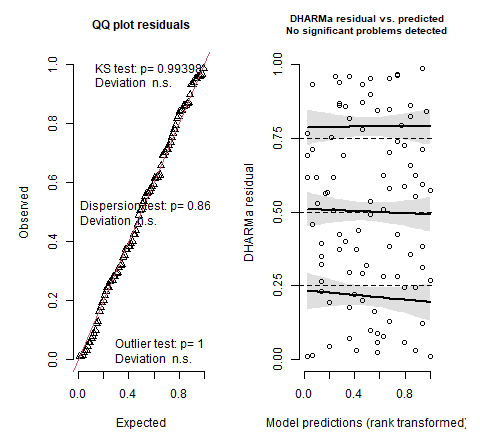}
    \caption{Simulated residual diagnostic plots for the negative binomial GLMM (collisions), generated using DHARMa to verify model assumptions and optimal fit.}
    \label{fig:collision_dharma}
\end{figure}

\begin{table}[t]
\centering
\caption{Descriptive statistics for collision frequency by navigational mode.}
\label{tab:collision_analysis}
\begin{tabular}{l S[table-format=1.2] S[table-format=1.1] S[table-format=1.0] S[table-format=2.0] S[table-format=1.0]}
\toprule
\textbf{Navigation Mode} & {\textbf{Mean}} & {\textbf{Median}} & {\textbf{Min}} & {\textbf{Max}} & {\textbf{Mode}} \\
\midrule
Normal Vision & 1.00 & 0.0 & 0 & 6 & 0 \\
Existing Aid & 1.96 & 0.0 & 0 & 15 & 0 \\
Proposed K1 & 4.39 & 2.0 & 0 & 35 & 0 \\
Proposed K3 & 3.60 & 1.5 & 0 & 17 & 1 \\
\bottomrule
\end{tabular}
\addvspace{5pt}
\small
\end{table}
\indent  The collision count was modeled so that it depends on the navigational mode, the trial index (first or second), and the variation of the scenario navigated by the subject, with a random intercept included for each subject to account for the design of repeated measures. To confirm that the optimal combination of fixed and random effects was selected for the negative binomial GLMM, the model's structural assumptions were  evaluated using the DHARMa package in R. The simulated residual diagnostics verify the goodness-of-fit and the robustness of this specific formula, as illustrated in Fig\ref{fig:collision_dharma}.
Using the Bonferroni method for $p$-adjustment, it was found that the random effect for the subject $(1 | \text{Subject})$ exhibited a variance of $\sigma^2 = 0.2961$ (SD = 0.5442 collisions). The collisions in the existing mode and proposed solutions were significant compared with the normal mode (as shown in Table \ref{tab:fixed_effects}). The incidence rate ratios detailed in Table \ref{tab:irr_results} show that collisions in the existing mode increased 4.27 times over normal vision, while the proposed Level 1 and Level 2 are 5.32 and 10.99 times the normal vision, respectively.\\

\begin{table}[!ht]
\centering
\caption{Fixed effects estimates: conditional model for collision frequency.}
\label{tab:fixed_effects}
\setlength{\tabcolsep}{4pt}
\begin{tabular}{l S[table-format=-1.4] S[table-format=1.4] S[table-format=-1.3] S[table-format=1.6]}
\toprule
\textbf{Variable} & {\textbf{Estimate ($\beta$)}} & {\textbf{Std. Error}} & {\textbf{$z$-value}} & {\textbf{$Pr(>|z|)$}} \\
\midrule
(Intercept) & -1.1642 & 1.1708 & -0.994 & 0.320064 \\
Mode: Existing & 1.4544 & 0.5616 & 2.590 & 0.009603 \\
Mode: Proposed K1 & 1.6711 & 0.4718 & 3.542 & 0.000397 \\
Mode: Proposed K3 & 2.3947 & 0.7081 & 3.382 & 0.000720 \\
Trial Index & -0.6065 & 0.2629 & -2.307 & 0.021065 \\
\bottomrule
\end{tabular}
\end{table}
\begin{table}[!ht]
\centering
\caption{Incidence rate ratios (IRR) for collision frequency (Bonferroni adjusted).}
\label{tab:irr_results}
\begin{tabular}{l S[table-format=-1.3] S[table-format=1.3] S[table-format=1.4]}
\toprule
\textbf{Contrast} & {\textbf{Estimate ($\beta$)}} & {\textbf{IRR ($e^{\beta}$)}} & {\textbf{$p$-value}} \\
\midrule
Normal vs. Existing & -1.454 & 0.234 & 0.0473 \\
Normal vs. Proposed K1 & -1.671 & 0.188 & 0.0022 \\
Normal vs. Proposed K3 & -2.395 & 0.091 & 0.0040 \\
\midrule
Existing vs. Proposed K1 & -0.217 & 0.805 & 0.9586 \\
Existing vs. Proposed K3 & -0.940 & 0.391 & 0.5441 \\
Proposed K1 vs. Proposed K3 & -0.724 & 0.485 & 0.6156 \\
\bottomrule
\end{tabular}
\end{table}

\FloatBarrier

\subsubsection{Navigational efficiency: Task completion duration}

The time taken to complete the navigation task was recorded in milliseconds (ms) then converted to seconds for better analysis. As summarized in Table \ref{tab:completion_time_stats}, The baseline (normal vision mode) had an average completion time of 33 seconds. The proposed Level 2 solution had the smallest average completion time. The existing-aid mode came next, with an average of 118.93 seconds. Finally, the proposed Level 2 solution had the longest average duration of 134.96 seconds.

\begin{figure}[htbp]
    \centering
    \includegraphics[width=0.5\linewidth]{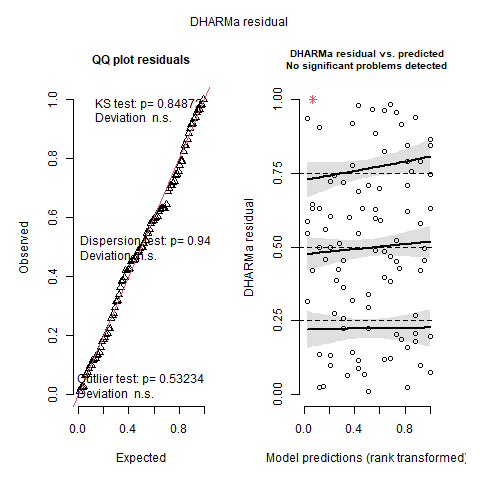}
    \caption{Simulated residual diagnostic plots for the Gamma GLMM (completion time), generated using DHARMa to verify model assumptions and optimal fit.}
    \label{fig:time_dharma}
\end{figure}

For time analysis, accounting for the non-negative, right-skewed nature of temporal data, a Gamma GLMM with log link was used using the formula:
\begin{equation}
    \text{Time} \sim \mathrm{Mode} + \mathrm{Scenario} + (1 \mid \mathrm{Subject})
\end{equation}
and the Bonferroni method for the p-adjustment.The completion time was modeled such that it depends on the navigational mode and scenario variation, again utilizing a random intercept for each subject.
Similarly, for temporal analysis, the selection of fixed and random effects for the gamma GLMM was validated via the DHARMa toolkit (see Fig \ref{fig:time_dharma}). The fixed effects estimates presented in Table \ref{tab:time_fixed_effects} show that all three modes were significantly ($p < 0.0001$) slower than normal. In particular, the existing-aid mode was 3.61 times slower than the baseline one and the proposed K1 solution was 3.82 times slower, while the proposed K3 solution was 3.03 times slower, i.e., the best among the compared modes. However, the differences between these modes were insignificant as shown in Table \ref{tab:time_irr_results}.
 The random subject variance was computed as $\sigma^2 = 0.0137$ (SD = 0.1171 seconds), indicating low temporal fluctuation between subjects across the combined trials.

\begin{table}[h]
\centering
\caption{Descriptive statistics for completion time (in seconds) by navigational mode.}
\label{tab:completion_time_stats}
\begin{tabular}{l S[table-format=3.2] S[table-format=3.2] S[table-format=2.2] S[table-format=3.2]}
\toprule
\textbf{Navigation Mode} & {\textbf{Mean}} & {\textbf{Median}} & {\textbf{Min}} & {\textbf{Max}} \\
\midrule
Normal Sighted    & 33.02  & 30.08  & 12.67 & 96.10  \\
Existing Aid      & 118.93 & 110.66 & 53.27 & 244.09 \\
Proposed K1       & 134.96 & 112.93 & 44.55 & 305.83 \\
Proposed K3       & 109.02 & 100.20 & 38.09 & 242.38 \\
\bottomrule
\end{tabular}
\end{table}

\begin{table}[htbp]
\centering
\caption{Fixed effects estimates: conditional model for completion time (Gamma GLMM).}
\label{tab:time_fixed_effects}
\setlength{\tabcolsep}{4pt}
\begin{tabular}{l S[table-format=1.4] S[table-format=1.5] S[table-format=2.3] S[table-format=1.2e-2]}
\toprule
\textbf{Variable} & {\textbf{Estimate ($\beta$)}} & {\textbf{Std. Error}} & {\textbf{$z$-value}} & {\textbf{$Pr(>|z|)$}} \\
\midrule
(Intercept) & 3.7867 & 0.28255 & 13.402 & {$< 0.0001$} \\
Mode: Existing & 1.2823 & 0.13497 & 9.501 & {$< 0.0001$} \\
Mode: Proposed K1 & 1.3386 & 0.12816 & 10.445 & {$< 0.0001$} \\
Mode: Proposed K3 & 1.1100 & 0.13458 & 8.248 & {$< 0.0001$} \\
\bottomrule
\end{tabular}
\end{table}

\begin{table}[htbp]
\centering
\caption{Incidence rate ratios (IRR) for completion time (Bonferroni adjusted).}
\label{tab:time_irr_results}
\begin{tabular}{l S[table-format=-1.4] S[table-format=1.3] S[table-format=1.4]}
\toprule
\textbf{Contrast} & {\textbf{Estimate ($\beta$)}} & {\textbf{IRR ($e^{\beta}$)}} & {\textbf{$p$-value}} \\
\midrule
Normal vs. Existing & -1.2823 & 0.277 & {$< 0.0001$} \\
Normal vs. Proposed K1 & -1.3386 & 0.262 & {$< 0.0001$} \\
Normal vs. Proposed K3 & -1.1100 & 0.330 & {$< 0.0001$} \\
\midrule
Existing vs. Proposed K1 & -0.0563 & 0.945 & 1.0000 \\
Existing vs. Proposed K3 & 0.1723 & 1.188 & 1.0000 \\
Proposed K1 vs. Proposed K3 & 0.2286 & 1.257 & 0.4552 \\
\bottomrule
\end{tabular}
\end{table}

\begin{table}[htbp]
\centering
\caption{Fixed effects estimates: steady-state model (trial 2 only) for completion time.}
\label{tab:time_steady_state}
\setlength{\tabcolsep}{4pt}
\begin{tabular}{l S[table-format=1.4] S[table-format=1.4] S[table-format=2.3] S[table-format=1.2e-2]}
\toprule
\textbf{Variable} & {\textbf{Estimate ($\beta$)}} & {\textbf{Std. Error}} & {\textbf{$z$-value}} & {\textbf{$Pr(>|z|)$}} \\
\midrule
(Intercept) & 3.6075 & 0.3379 & 10.676 & {< 2e-16} \\
Mode: Existing & 1.2704 & 0.1634 & 7.775 & {7.57e-15} \\
Mode: Proposed K1 & 1.2612 & 0.1617 & 7.802 & {6.10e-15} \\
Mode: Proposed K3 & 0.9605 & 0.1634 & 5.880 & {4.11e-09} \\
\bottomrule
\end{tabular}
\end{table}

Further analysis involved using only the second trials, discarding the first ones based on results of the Friedman test. Notably, the coefficient for the proposed Level 2 mode decreased from 1.11 to 0.9605. This mode still remains 2.6 times faster than the baseline mode, confirming its superiority as the fastest of the three modes. Concurrently, the random subject variance increased to $\sigma^2 = 0.0512$ (SD = 0.2263 seconds).


\subsection{Qualitative analysis of user feedback}

Although our quantitative analysis did not reveal large statistically significant differences between the conventional and proposed solutions in terms of collision counts and completion time, the qualitative findings suggest that the two approaches were perceived differently by users. Participants identified distinct advantages and disadvantages for each navigation mode. These contrasting strengths and weaknesses may have balanced each other numerically, resulting in no substantial statistical difference in performance metrics.

For the conventional mode, 8 positive and 11 negative coded responses were identified. Participants primarily valued object and depth identification (n = 6) and visual clarity (n = 2). However, the most frequent concerns were ambiguity of the scene (n = 7) and visual clutter (n = 4), suggesting that although object recognition was supported, overall spatial interpretation remained challenging. On the other hand, the proposed mode generated 13 positive and 13 negative coded responses. Its main strengths were direct pathfinding (n = 7) and cognitive simplicity (n = 6). Nevertheless, limitations included system instability (n = 6), misleading feedback (n = 5), and panic or increased cognitive load (n = 2). These contrasting advantages and disadvantages likely contributed to the absence of strong statistical differences in quantitative performance measures, while revealing important usability trade-offs between the two approaches.

\section{Discussion}

\subsection{Task success and metric validity}
\label{subsec:task_success}
Although the task success metric was not a discriminative metric to assess the differences between modes (with a 98.9\% success rate across trials), it provided critical evidence that the proposed solution is functional and successfully achieves the goal of helping visually impaired users reach their destination. Consequently, this high success rate guided our evaluation towards collision count and completion time to assess how well the solution worked, rather than if it just worked.

\subsection{Navigational collision safety vs. system stability}
\label{subsec:safety_stability}
Exhibiting high subject variance alongside a small sample size makes it difficult to draw definitive statistical conclusions, but the data suggests a valid performance parity. In terms of the collision metric, the proposed Level 1 (K1) solution performed closely to the existing conventional solution. The difference between these solutions had only $\beta = 0.2$ with an IRR of 0.8. Thus, this difference is numerically  in the range of 23--25\% from the existing solution ($p = 0.95$).

However, the qualitative data reveals that the solutions achieved these numbers in fundamentally different ways. The existing conventional solution achieved its safety performance $1.96 \pm 3.54 \text{ collisions}$ by providing users with images representing what they were actually seeing, allowing them to navigate by mapping low-resolution images to known objects. In contrast, the proposed solutions relieved this recognition load by providing assisted pathfinding. This shifted the cognitive burden away from the user, as they only needed to know \textit{where} to walk, rather than \textit{what} the obstacle was.

On the other hand, system instability in the proposed modes caused some users to experience more collisions. This reinforces the point regarding visual clutter: when the proposed solution was used with higher object density (Level 2), collision rates increased. A critical qualitative insight revealed that when an obstacle came very close to the user, the entire field of view (FOV) was activated, showing only white maps. This sudden overwhelming feedback caused some subjects to panic, impairing their spatial decision-making and leading to compounding collisions.

\subsection{Temporal efficiency}
\label{subsec:temporal_efficiency}
The task completion times demonstrated very close results across the sample means. The GLMM outcomes estimate that the proposed K1 and K3 modes took 1.058 and 0.841 times the duration of the existing solution, respectively. While these modes were significantly slower than the normal no-device baseline, the differences in completion time between the electronic modes themselves were statistically insignificant.

By examining the qualitative data, we suggest that the assisted pathfinding ultimately helped users navigate more efficiently. Furthermore, since the proposed K3 mode was always tested after the K1 mode, this fixed trial order likely provided users with a better understanding of the system's logic. As the novelty wore off, users learned to trust the pathfinding feedback, resulting in faster completion times.

Finally, the quantitative results suggest that the proposed Level 1 mode matched closely the existing conventional mode in both navigational safety and temporal efficiency. However, due to the high individual variance and the small sample size, these observations cannot be definitively confirmed. The high p-values (e.g., $p = 0.95$ for K1 versus the conventional solution) leave a critical statistical question open: we cannot be certain if the p-value is large because the two modes share true performance parity, or simply because the study lacked the statistical power to detect the true effect through the noise of human variability. Ultimately, the qualitative data provides the most important insight, proving that the conventional and proposed solutions place fundamentally different cognitive demands on the user.

\subsection{Experimental limitations}
\label{subsec:limitations}

Although our framework offers promising results and insights, we acknowledge that it still has some certain limitations which we enlist and describe briefly as follows.

\subsubsection{Sample size} The study's sample size may have underpowered the statistical analysis, potentially limiting the ability to detect smaller significant effects or generalized trends across a broader population.

\subsubsection{Wireless connectivity} The reliance on a wireless connection between the mobile device and the laptop (where the core processing occurs) may have introduced latency or stability issues, potentially affecting real-time performance and data consistency.  

\subsubsection{Lack of environment dynamics} The experiments were conducted using only static (stationary) objects. The system was not tested with moving obstacles or objects that emit sound. This restricted setup may not fully represent the complexity of a real-world dynamic environment. 

\subsubsection{Lack of eye-tracking} Because the system lacked active eye-tracking, the field of view was set to a fixed value of 17° to cover the potential range of a user's gaze. While the PRIMA implant itself is smaller (with around 7° FOV), a larger area was projected to ensure the implant could still receive information as the user moved their eyes. 

\subsubsection{Lack of temporal perceptual modeling} Temporal aspects were not accounted for in the proposed spatial perceptual model. Consequently, the temporal stimulation effects, such as the fading effect, were not taken into consideration.

\section{Conclusion and Future Work}
In conclusion, this study addresses the challenges of the existing visual prosthetics, primarily the severe resolution constraints. Given the historical lack of user-centered design of such prosthetics, we first conducted interviews with a group of visually impaired people to know their actual needs as users of visual prosthetics. Thematic analysis of the outcomes of these interviews revealed that independent navigation and collision avoidance, rather than detailed scene comprehension, are the most critical daily challenges. To address these challenges, we developed an end-to-end pipeline tailored for outdoor navigation of visually impaired pedestrians. Instead of trying to imitate natural vision by rendering raw, cluttered scenes, we adopted a “smart approach” where visual prostheses are regarded as an assistive technology designed specifically to restore functional, task-specific vision. A central piece of this system is our priority-scoring translator, which strictly considers the resolution limits of the prostheses to filter out visual clutter and show only what is essential for safe navigation, such as immediate obstacles and clear paths. Experimental validation of this pipeline yielded a high task success rate of 98.9\%. While quantitative results showed comparable performance to conventional aids in terms of temporal efficiency and collision safety, qualitative feedback highlighted a fundamental shift in the user experience. By translating complex environments into simple geometric primitives, the system significantly reduced the cognitive burden of object recognition. Ultimately, users no longer needed to figure out what an obstacle was; they simply needed to know where it was to navigate safely. This work presents a significant step toward making bionic vision aligned with the practical needs of the visually impaired individuals within the constraints of current prostheses. \\
\indent While the proposed framework demonstrates a significant step toward smart, task-aware vision, there are still important enhancements planned for future work. First, the proposed  pipeline should be extended to handle dynamic environments. Also, our current object priority scoring module relies primarily on object depth. This module could be eventually enriched to incorporate additional factors, such as the semantic risk factor (i.e., cars have higher priority than garbage bins) and the velocity of the moving objects, to create a more robust scene filtering system. Furthermore, to better simulate real-world conditions and enhance the ecological validity of our framework, we intend to move our testing environment from virtual reality (VR) to mixed reality (MR). This extension will minimize the gap between simulated environments and actual physical environments, which are usually more difficult to navigate.

\section*{\LARGE Declarations}
\subsection*{Funding}
The authors would like to acknowledge financial support received from the Information Technology Development Agency (ITIDA), the Academy of Scientific Research \& Technology (ASRT) in Egypt, and Misr El Kheir.

\subsection*{Competing Interests}
The authors have no relevant financial or non-financial interests to disclose. 

\subsection*{Data Availability}
The data can't be shared publicly for matters of privacy. 

\subsection*{Ethical Approval}
This study was conducted in accordance with established ethical standards and approved by the relevant institutional review board at Cairo University. All procedures involving human participants were performed in compliance with the ethical principles outlined in the Declaration of Helsinki and its later amendments.

\subsection*{Consent to Participate}
 Informed consent was obtained from all individual participants included in the study prior to their participation. Confidentiality and anonymity of participant data were strictly maintained throughout the research process.

\subsection*{Consent to Publish}
Not applicable.

\subsection*{Acknowledgment}
The authors would like to thank Dr. M. Hassib (Fortiss Research Institute in Munich) and Dialogue in the Dark - Egypt for their invaluable guidance and facilitation of the user interviews. We also appreciate Dr. J. Granley (UCSB) for his insights on PRIMA implant perception, Prof. A. Rehan (Cairo University) for statistical analysis support, and all volunteer participants.

\bibliographystyle{ieeetr}
\bibliography{refs}

\end{document}